\documentclass[]{pasj02}
\usepackage[switch,mathlines]{lineno} 
\usepackage{xspace}
\usepackage{url}
\usepackage{comment}
\newcommand{\nar}{New Astronomy Reviews}
\newcommand{\gmas}{g_{\rm m}}
\newcommand{\fmas}{f_{\rm m}}
\newcommand{\pllx}{\varpi}
\newcommand{\ebv}{{\rm E(B-V)}}
\newcommand{\Av}{{\rm A_{\rm V}}}
\newcommand{\Rv}{{\rm R_{\rm V}}}
\newcommand{\fmassin}{f_{\rm m} \sin^{-3} i_\mathrm{iso}}

\newcommand{\kvisi}{K_1}
\newcommand{\kbase}{29.8}
\newcommand{\argpr}{\omega_1}
\newcommand{\pout}{P_{\rm out}}
\newcommand{\eout}{e_{\rm out}}
\newcommand{\iout}{i_{\rm out}}
\newcommand{\tprigaia}{t_{\rm p,out,2016}}
\newcommand{\tprimine}{t_{\rm p,out,2023.5}}
\newcommand{\vcom}{\gamma}
\newcommand{\pin}{P_{\rm in}}

\newcommand{\fehn}{{\rm [Fe/H]}}
\newcommand{\fehi}{{\rm [Fe/H]_{\rm init}}}

\newcommand{\teffp}{T_{\rm eff,1}}
\newcommand{\teffs}{T_{\rm eff,2}}
\newcommand{\tefft}{T_{\rm eff,3}}

\newcommand{\massp}{m_{\rm 1}}
\newcommand{\masss}{m_{\rm 2}}
\newcommand{\masst}{m_{\rm 3}}

\newcommand{\radip}{R_1}
\newcommand{\radis}{R_2}
\newcommand{\radit}{R_3}
\newcommand{\tprim}{G1010\xspace}

\newcommand{\msun}{{\rm M}_\odot}
\newcommand{\rsun}{{\rm R}_\odot}

\newcommand{\tage}{t_{\rm age}}

\jyear{2025}
\Received{}
\Accepted{}

\begin{document} 

\title{Discovery of a compact hierarchical triple main-sequence star
  system while searching for binary stars with compact objects}

\author{
 Ataru \textsc{Tanikawa},\altaffilmark{1}\orcid{0000-0002-8461-5517}\altemailmark\email{tanik@g.fpu.ac.jp} 
 Akito \textsc{Tajitsu},\altaffilmark{2}\orcid{0000-0001-8813-9338}
 Satoshi \textsc{Honda},\altaffilmark{3}\orcid{0000-0001-6653-8741}
 Hiroyuki \textsc{Maehara},\altaffilmark{2}\orcid{0000-0003-0332-0811}
 Bunei \textsc{Sato},\altaffilmark{4}\orcid{0000-0001-8033-5633}
 Kento \textsc{Masuda},\altaffilmark{5}\orcid{0000-0003-1298-9699}
 Masashi \textsc{Omiya},\altaffilmark{6}\orcid{0000-0002-5051-6027}
 Hideyuki \textsc{Izumiura},\altaffilmark{2}
}
\altaffiltext{1}{Center for Information Science, Fukui Prefectural
  University, 4-1-1 Matsuoka Kenjojima, Eiheiji-cho, Fukui 910-1195,
  Japan}
\altaffiltext{2}{Subaru Telescope Okayama Branch, National
  Astronomical Observatory of Japan, 3037-5, Honjou, Kamogata,
  Asakuchi, Okayama 719-0232, Japan}
\altaffiltext{3}{Nishi-Harima Astronomical Observatory, Center for
  Astronomy, University of Hyogo, Sayo, Hyogo 679-5313, Japan}
\altaffiltext{4}{Department of Earth and Planetary Sciences, Institute
  of Science Tokyo, 2-12-1 Ookayama, Meguro-ku, Tokyo 152-8551, Japan}
\altaffiltext{5}{Department of Earth and Space Science, Graduate
  School of Science, Osaka University, 1-1 Machikaneyama-cho,
  Toyonaka, Osaka 560-0043, Japan}
\altaffiltext{6}{Astrobiology Center, 2-21-1 Osawa, Mitaka, Tokyo
  181-8588, Japan}




\maketitle

\begin{abstract}
  We have discovered a compact hierarchical triple main-sequence star
system, which is cataloged as Gaia DR3 1010268155897156864 or TIC
21502513. Hereafter, we call it ``\tprim''. \tprim consists of a
primary (the most massive) star and inner binary that orbit each
other. The primary star is a $0.85_{-0.03}^{+0.03}\;\msun$
main-sequence (MS) star, and the inner binary components are
$0.63_{-0.02}^{+0.02}$ and $0.61_{-0.02}^{+0.02}\;\msun$ MS stars. The
outer and inner orbital periods are $277.2_{-1.3}^{+1.6}$ and $\sim
18.26$ days, respectively. \tprim is categorized as a single-lined
spectroscopic binary, and its orbital solution indicates that \tprim
possibly accompanies a massive compact object, such as a neutron star
or massive white dwarf. In order to confirm the presence of a massive
compact object, we have performed several-times low signal-to-ratio
(SNR) and one-time high SNR spectroscopic observations, and determined
the outer orbital parameters. Moreover, we have deeply analyzed the
high SNR spectroscopic data, and found that \tprim accompanies not a
massive compact object, but an inner binary. We have investigated
\tprim's light curve in Transiting Exoplanet Survey Satellite
(TESS), and concluded that the inner binary is actually an eclipsing
binary, not included in TESS Eclipsing Binary Stars. We have obtained
the inner orbital parameters from the TESS light curve. \tprim is
similar to compact hierarchical triple star systems previously
discovered by eclipse timing variation analysis. Our discovery has
shown that such triple star systems can be discovered by combination
of low- and high-SNR spectroscopic observations with the help of Gaia
DR3 and the upcoming Gaia DR4/DR5.

\end{abstract}



\section{Introduction}
\label{sec:Introduction}

White dwarfs (WDs), neutron stars (NSs), and black holes (BHs) are
compact objects that stars leave behind after they exhaust their
nuclear fuels. Isolated compact objects simply cools and fades over
time. However, compact objects often merge, and their mergers may
cause fascinating astronomical transients: BH-BH and BH-NS mergers as
gravitational wave sources \citep{2023PhRvX..13d1039A}, NS-NS mergers
as multi-messenger sources \citep{2017ApJ...848L..12A}, and WD-WD
mergers as one of the most promising progenitors for type Ia
supernovae \citep{1984ApJS...54..335I, 1984ApJ...277..355W}. This
motivates astronomers to search for binary stars with compact objects,
hereafter compact binaries.

Very recently, compact binaries have been discovered from Gaia Data
Release 3 (DR3: \cite{2023A&A...674A..34G}) with spectroscopic
follow-up observations, such as Gaia BHs (\cite{2023MNRAS.518.1057E};
\yearcite{2023MNRAS.521.4323E}; \cite{2023ApJ...946...79T};
\cite{2023AJ....166....6C}; \cite{2024A&A...686L...2G}), Gaia NSs
(\cite{2024OJAp....7E..38E}; \yearcite{2024OJAp....7E..58E}), detached
WD binaries (\cite{2024MNRAS.527.11719};
\yearcite{2024PASP..136h4202Y}; \cite{2024ApJ...964..101Z}), and a
binary consisting of hot subdwarf star and compact object
\citep{2023A&A...677A..11G}. These compact binaries have orbital
periods of $\sim 10^2$ -- $10^3$ days, which are much longer than
compact binaries previously discovered by X-ray and radio observations
\citep{2024NewAR..9801694E}.

However, there remains a possibility that compact objects in these
binaries are indeed two objects. In other words, these compact
binaries could be triple star systems. Hereafter, we call these
compact objects and compact binaries ``compact object candidates'' and
``compact binary candidates'', respectively. The compact object
candidates of Gaia BH2 and BH3 may be inner double BHs instead of
single BHs, while \citet{2024PASP..136a4202N} have ruled out the
possibility that Gaia BH1 contains inner double BHs. Note that
\citet{2025OJAp....8E..79T} have pointed out the formation rate of
such triple systems is not rare if they are formed in open
clusters. The possibility that the compact object candidates of Gaia
NSs are inner double WDs has not yet been excluded
(\cite{2024OJAp....7E..38E}; \cite{2024OJAp....7E..58E}). Detached WD
binaries may not be binaries with WDs and main-sequence (MS) stars,
but triple MS stars (\cite{2024MNRAS.527.11719};
\yearcite{2024PASP..136h4202Y}). One reason why there remain these
possibilities is that these compact binary candidates have long
periods. They can be stable even if they contain inner binaries. Note
that many compact binaries with orbital periods of $\lesssim 10$ days
have been also discovered recently (\cite{2022MNRAS.517.4005M};
\cite{2022ApJ...936...33Z}; \yearcite{2023SCPMA..6629512Z};
\cite{2022NatAs...6.1085S}; \cite{2022ApJ...938...78L};
\cite{2022ApJ...940..165Y}; \cite{2022NatAs...6.1203Y};
\cite{2023AJ....165..187Q}; \cite{2024AJ....168..217D};
\cite{2024MNRAS.529..587R}; \cite{2024ApJ...961L..48Z};
\cite{2025OJAp....8E..61T}; \cite{2025JHEAp..45..428Z}
\cite{2025arXiv250912808S}). They are more unlikely to contain inner
binaries because of their short orbital periods.

In this paper, we report our survey and discovery of a triple star
system during the search for compact binaries with orbital periods of
$10^2$ -- $10^3$ days. Because we search for such compact binaries,
our triple star system has outer orbital periods of $10^2$ -- $10^3$
days. Such triple star systems are categorized into ``compact'' triple
star systems because of their short orbital periods. Previous studies
have discovered similar triple star systems via eclipsing time
variation analysis (\cite{2011MNRAS.417L..31S};
\cite{2012AJ....143..137G}; \yearcite{2015AJ....150..178G};
\cite{2013ApJ...768...33R}; \yearcite{2022MNRAS.513.4341R};
\yearcite{2024A&A...686A..27R}; \cite{2013MNRAS.428.1656B};
\yearcite{2015MNRAS.448..946B}; \yearcite{2016MNRAS.455.4136B};
\yearcite{2020MNRAS.493.5005B}; \yearcite{2025A&A...695A.209B};
\yearcite{2025arXiv251004565B}; \cite{2013ApJ...763...74L};
\yearcite{2014AJ....148...37L}; \yearcite{2015AJ....149...93L};
\cite{2014AJ....147...86T}; \yearcite{2014AJ....147...87T};
\cite{2014AJ....147...45C}; \cite{2015AJ....149..197Z};
\cite{2015A&A...577A.146B}; \cite{2019MNRAS.485.2562H};
\yearcite{2022MNRAS.509..246H}; \cite{2020MNRAS.498.6034M};
\yearcite{2024A&A...685A..43M}; \cite{2022A&A...668A.173G};
\cite{2022ApJ...924...66Y}; \cite{2022MNRAS.511.4710E};
\cite{2023MNRAS.521.1908M}; \yearcite{2024A&A...690A.153M};
\cite{2023MNRAS.526.2830C}; \cite{2023MNRAS.522.3076Y};
\cite{2024ApJ...974...25K}; \cite{2025arXiv251105761R}). Our triple
star system is not included in their discoveries. It is not found in a
list of quadruple star candidates \citep{2026AJ....171...29K}.
Moreover, we discover it by only spectroscopic observations. Indeed,
after the discovery, we recognize that its inner binary is an
eclipsing binary from the light curve of Transiting Exoplanet Survey
Satellite (TESS). The delay in our recognition results from the fact
that the triple star system is not included in catalogs of eclipsing
binaries, such as TESS Eclipsing Binary Stars
(\cite{2022ApJS..258...16P}).

The remainder of this paper is structured as follows. In section
\ref{sec:Terminology}, we organize our terminology, and show the
parameters of our triple star system in advance. In section
\ref{sec:Selection}, we describe how to select our targets from Gaia
DR3. In section \ref{sec:Follow-up}, we introduce the instruments we
use for our follow-up observations. In section \ref{sec:Outer_orbit},
we present our analysis to determine the outer orbit of our triple
star system. In sections \ref{sec:sb3} and
\ref{sec:Inner_orbit}, we show how to detect the inner binary, and
determine its orbital parameters. In section \ref{sec:Comparison}, we
compare our triple star system with triple star systems discovered by
previous surveys. In section \ref{sec:Conclusions}, we makes
conclusions in this paper.

\section{Terminology and notations}
\label{sec:Terminology}

\begin{figure}
 \begin{center}
   \includegraphics[width=8cm]{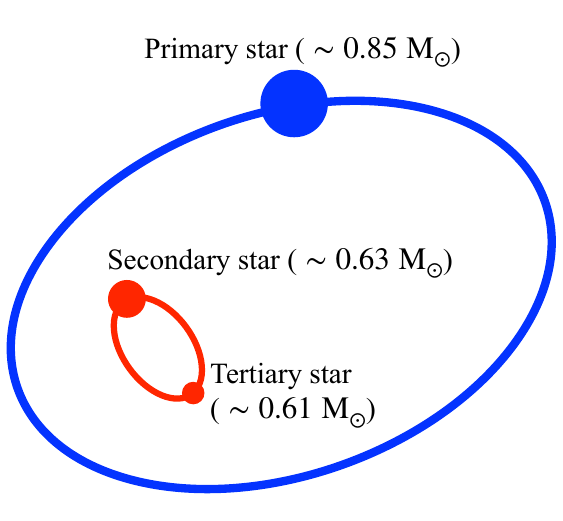}
 \end{center}
 \caption{Primary, secondary, and tertiary stars of our triple stars.}
\label{fig:terminology}
\end{figure}

We organize our terminology in this paper. Figure
\ref{fig:terminology} shows what stars are called ``primary'',
``secondary'', and ``tertiary'' stars. We name them in order of
decreasing mass and optical luminosity. Note that our naming rule
might be different from the naming rule of researchers who discover
triple star systems via eclipsing time variation analysis. They name
components of a triple star system in order from the inside out. In
other words, they call our primary star ``tertiary star''. This
difference comes from the fact that we first recognize the presence of
the primary star, while they finally recognize it.

We call the pair of the secondary and tertiary stars ``inner binary'',
or ``faint companion'' of the primary star. We use the term of the
faint companion, when we analyze the radial velocity (RV) variation of
the primary star (see sections \ref{sec:Selection} and
\ref{sec:Outer_orbit}). In the analysis, we just regard it as a single
object. We call the orbit of the inner binary ``inner orbit'', and the
relative motion between the primary star and inner binary ``outer
orbit''.

We summarize the parameters of Gaia DR3 1010268155897156864 or TIC
21502513 in Table \ref{tab:Summary}. We abbreviate it to \tprim. We
divide these parameters into four: the parameters provided by Gaia
DR3, the outer orbit parameters derived by our RV variation analysis,
the parameters of the three stars obtained from high
signal-to-noise-ratio (SNR) spectroscopy, and the inner orbit
parameters estimated from their eclipses. We attach the subscriptions
``out'' and ``in'' to parameters related to the outer and inner
orbits, respectively. We also attach the subscriptions ``1'', ``2'',
and ``3'' to parameters related to the primary, secondary, and
tertiary stars, respectively.

\begin{table}
  \tbl{Summary of Gaia DR3 1010268155897156864 or TIC 21502513.}{%
    \begin{tabular}{ll}
      \hline
      {\bf Names} \\
      Abbreviation & \tprim \\
      Gaia source ID & 1010268155897156864 \\
      TESS Identity Catalog & 21502513 \\
      {\bf Parameters from Gaia DR3} \\
      NSS solution type & SB1 \\
      Right Ascension (RA) & \timeform{8h51m50.60s} \\
      Declination (Dec)    & \timeform{44D55'11.91''} \\
      Apparent magnitude in G band (G) [mag] & $12.67$ \\
      Renormalized Unit Weight Error (RUWE)  & $4.58$  \\
      Parallax ($\pllx$) [mas] & $2.30 \pm 0.25$\footnotemark[$*$] \\
      \multicolumn{2}{l}{{\bf Outer orbit from RV variation}} \\
      RV semi-amplitude ($\kvisi$) [km/s]       & $25.36_{-0.66}^{+0.69}$ \\
      Orbital period ($\pout$) [day]            & $277.2_{-1.3}^{+1.6}$ \\
      Orbital eccentricity ($\eout$)            & $0.230_{-0.018}^{+0.019}$ \\
      Argument of periastron ($\argpr$) [deg]   & $166.2_{-6.2}^{+6.0}$ \\
      Periastron passage ($\tprigaia$) [day]    & $119.7_{-18.9}^{+10.1}$ \\
      Periastron passage ($\tprimine$) [day]    & $-122.0_{-2.9}^{+3.2}$ \\
      Center-of-mass velocity ($\gamma$) [km/s] & $-4.25_{-0.33}^{+0.34}$ \\
      Mass function ($\fmas$) [$\msun$]         & $0.431_{-0.031}^{+0.033}$ \\
      \multicolumn{2}{l}{{\bf SB3 spectral and isochrone fittings}} \\
      Age ($\tage$) [Gyr]                   & $12.17_{-1.87}^{+1.16}$ \\
      Initial metallicity ($\fehi$)         & $-0.33_{-0.09}^{+0.09}$ \\
      Effective temperature ($\teffp$) [K]  & $5869_{-60}^{+65}$    \\
      Mass ($\massp$) [$\msun$]             & $0.85_{-0.03}^{+0.03}$  \\
      Radius ($\radip$) [$\rsun$]           & $1.10_{-0.07}^{+0.08}$  \\
      Effective temperature ($\teffs$) [K]  & $4584_{-71}^{+75}$ \\
      Mass ($\masss$) [$\msun$]             & $0.63_{-0.02}^{+0.02}$ \\
      Radius ($\radis$) [$\rsun$]           & $0.63_{-0.02}^{+0.02}$ \\
      Effective temperature ($\tefft$]) [K] & $4454_{-67}^{+70}$ \\
      Mass ($\masst$) [$\msun$]             & $0.61_{-0.02}^{+0.02}$ \\
      Radius ($\radit$) [$\rsun$]           & $0.61_{-0.02}^{+0.02}$ \\
      Center-of-mass velocity ($\gamma$) [km/s]              & $-4.93^{+0.18}_{-0.16}$ \\
      Modified mass function ($\fmas \sin^{-3} i$) [$\msun$] & $0.432^{+0.016}_{-0.017}$ \\
      \multicolumn{2}{l}{{\bf Inner orbit from eclipses}} \\
      Orbital period ($\pin$) [day] & $18.3$\footnotemark[$\dag$]\\
      \hline
  \end{tabular}} \label{tab:Summary}
  \begin{tabnote}
    \footnotemark[$*$] The uncertainty of parallax is inflated, based
    on the formula proposed by \citet{2025OJAp....8E..62E}. The
    inflation factor is $3.34$ for \tprim. \\
    \footnotemark[$\dag$] We do not show the uncertainty of the inner
    orbital period, because we do not rigorously analyze eclipse
    timing.\\
  \end{tabnote}
\end{table}

\section{Selection of targets}
\label{sec:Selection}

Although we report the discovery of one triple star system in this
paper, we primarily search for detached compact binaries from the
table ``\verb|nss_two_body_orbit|'' in Gaia DR3
\citep{2023A&A...674A..34G}. Here, we describe our method to narrow
down compact binary candidates. The table \verb|nss_two_body_orbit|
contains about 200,000 objects with binary orbital solutions. They are
categorized into ``\verb|Orbital|'', ``\verb|SB1|'', and
``\verb|AstroSpectroSB1|'' by a data type
``\verb|nss_solution_type|'', where the binary orbital solution of an
object with ``\verb|Orbital|'', ``\verb|SB1|'', and
``\verb|AstroSpectroSB1|'' is based on astrometric data, spectroscopic
data, and both astrometric and spectroscopic data, respectively. We
select our targets, put the following constraints on these objects.
\begin{enumerate}
\item The apparent magnitude of an object is less than $13$ mag in the
  G band because of the limiting magnitude of the instruments we use.
\item The declination of an object is more than $-20$ degree because
  of our observing sites.
\item The primary star of an object is G- or K-type star. Such a star
  has many absorption lines in the optical band.
\item The primary star of an object has a faint companion whose mass
  is larger than, or equal to $1.35\;\msun$ (described in detail
  below). \label{item:darkmass}
\item The primary star of an object does not outshine its faint
  companion if the faint companion is a single star in its zero-age
  (ZA) main-sequence (MS) phase (described in detail
  below). \label{item:visiblemass}
\end{enumerate}

In the constraint \ref{item:darkmass}, we estimate the faint companion
mass for \verb|Orbital| and \verb|AstroSpectroSB1|, or the minimum
mass of the faint companion for \verb|SB1|. For this purpose, we need
the primary star mass in addition to the binary orbital solution. We
obtain the primary star mass as follows.  We construct a table
consisting of ZAMS star masses and their G-band absolute magnitude,
based on the MESA Isochrones \& Stellar Tracks (MIST) model
(\cite{Paxton11}; \yearcite{2013ApJS..208....4P};
\yearcite{2015ApJS..220...15P}; \cite{2016ApJ...823..102C};
\cite{2016ApJS..222....8D}), where we adopt the solar metallicity. We
derive the primary star mass, giving the table its G-band absolute
magnitude as input. Indeed, we do not estimate the primary star mass
accurately. The primary star does not always have the solar
metallicity, and is not a ZAMS star. Moreover, we do not take into
account the interstellar extinction of the primary star. We use the
estimated mass just as a simple indicator in order to deal with a
large number of objects. If the G-band luminosity of a primary star is
fixed, the primary star tends to have small masses as its age
increases, and its metallicity decreases. In other words, we would
overestimate the primary star masses overall, which effectively means
we overestimate the faint companion masses. Consequently, our selected
targets can have faint companions of $< 1.35\;\msun$.

The constraint \ref{item:visiblemass} enables us to avoid an object
which contains a faint companion difficult to identify as either
compact object or non-degenerate star like a MS star. For example,
even if the faint companion mass is $2\;\msun$, we cannot identify if
it is a NS or just a MS star, when the primary star is much more
luminous than a $2\;\msun$ MS star. For this constraint, we estimate
the G-band absolute magnitude of a faint companion, referring to the
table constructed above, given that the faint companion is a ZAMS
star. When its G-band absolute magnitude is more than that of the
primary star, we remove the object from our targets.

We obtain a target list containing $\sim 50$ objects as detached
compact binary candidates. From our target list, we focus on one
candidate, which is identified as a triple star system later. We name
it \tprim after the initial character of ``Gaia'' and the first four
digits of it Gaia source ID. \tprim is the first to have its identity
determined in our target list. We perform follow-up observations for
several objects, however have not yet found out what they are.

\section{Follow-up observations}
\label{sec:Follow-up}

\subsection{Gunma Astronomical Observatory Echelle Spectrograph for Radial Velocity}
\label{sec:GAOES-RV}

We obtain 17 spectra with the Gunma Astronomical Observatory Echelle
Spectrograph for Radial Velocity (GAOES-RV,
\cite{2024SPIE13096E..44S}) on the 3.8-m Seimei telescope at the
Okayama observatory from the second half of 2023 to the first half of
2025 (programmes 23B-N-CN02, 24A-N-CN03, 24A-K-0024, 24B-N-CN04,
24B-K-0013, 25A-N-CN18, and 25A-K-0023). The wavelength coverage is
516 -- 593 nm. The spectral resolution is $R \sim 65000$. We use the
ThAr lamp as a wavelength calibration light source, not using the
iodine (I$_2$) cell. This is because we do not need an accuracy of
$\sim 10$ m/s, and the apparent magnitudes of our targets are close to
the limiting magnitude of GAOES-RV. Exposure times range from 900 to
1800 s for each spectrum, depending on weather conditions. The SNR per
pixel is typically $\sim 20$.

\subsection{Medium And Low-dispersion Long-slit Spectrograph}
\label{sec:MALLS}

We obtain 9 spectra with the Medium And Low-dispersion Long-slit
Spectrograph (MALLS) on the 2-m NAYUTA Telescope at the Nishi-Harima
Astronomical Observatory from the second half of 2023 to the first
half of 2025. We adopt the \timeform{1.2''}-width slit and the 1800
l/mm grating. This settings achieve a wavelength coverage 40 nm
centered on the wavelength range of H$_\alpha$. The spectral
resolution is $R \sim 7500$. Exposure times are $3000$ s. The SNR per
pixel is typically $\sim 20$.

\subsection{High Dispersion Spectrograph}
\label{sec:HDS}

We obtain 1 spectra with the High Dispersion Spectrograph (HDS) on the
8.2-m Subaru telescope at the NAOJ Hawaii Observatory. The settings
are as follows. We choose the standard spectrograph setup ``StdYb'' in
which the wavelength regions are 414 -- 535 nm and 559 -- 681 nm. The
slit width we use is \timeform{0.4''}, which provides a resolving
power of $R \sim 90~000$. Exposure times are 1500 s for each
spectrum. The SNR per pixel is typically $\sim 200$.

\section{Outer orbit from radial velocity variation}
\label{sec:Outer_orbit}

We summarize the RVs of the primary star in \tprim in Table
\ref{tab:RV_g1010}. We obtain a RV from each spectrum as follows. For
GAOES-RV data, we identify $60$ -- $70$ absorption lines between 517
and 533 nm in each spectrum. We fit the absorption lines with a
Gaussian, compare their median values with the Atomic Spectra Database
(ASD) provided by National Institute of Standards and
Technology\footnote{\url{https://www.nist.gov/pml/atomic-spectra-database}},
and derive a RV for each absorption line. We take the arithmetic mean
of all these RVs, and regard it as the representative value at the
epoch. We regard the standard error of all these RVs as the error of
the representative value of the RV. For the HDS data, we adopt the
same method to obtain the RV, except that we find $100$ absorption
lines between 517 and 534 nm in the HDS spectrum. For MALLS data, we
spot $5$ absorption lines in each spectrum. We obtain RVs in the same
way as for the case of GAOES-RV, except that we refer to the atomic
line list in \citet{1995all..book.....K}. The atomic line database is
different from those referred in the case of GAOES-RV and HDS, however
the resulting systematic error in RVs should be $\lesssim 0.1$
km/s. It does not affect our results.

\begin{table}
  \tbl{RVs of the primary star in \tprim.}{%
    \begin{tabular}{ccc}
      \hline
      Julian epoch year & RV [km/s] & Instrument  \\ 
      \hline
      $2023.8502$ & $ -19.3 \pm  0.8$ & GAOES-RV \\
      $2023.9325$ & $ -34.7 \pm  1.0$ & GAOES-RV \\
      $2023.9352$ & $ -35.8 \pm  1.2$ & GAOES-RV \\
      $2023.9920$ & $ -29.8 \pm  1.3$ & GAOES-RV \\
      $2024.0109$ & $ -26.6 \pm  1.5$ & GAOES-RV \\
      $2024.0275$ & $ -22.7 \pm  1.6$ & GAOES-RV \\
      $2024.0301$ & $ -22.2 \pm  1.3$ & GAOES-RV \\
      $2024.0414$ & $ -22.9 \pm  2.4$ & MALLS \\
      $2024.0797$ & $  -8.5 \pm  5.3$ & MALLS \\
      $2024.0876$ & $ -10.0 \pm  1.3$ & GAOES-RV \\
      $2024.0989$ & $  -8.9 \pm  4.1$ & MALLS \\
      $2024.1287$ & $  -2.5 \pm  0.9$ & GAOES-RV \\
      $2024.1509$ & $  -1.1 \pm  4.9$ & MALLS \\
      $2024.1890$ & $  +5.8 \pm  1.0$ & GAOES-RV \\
      $2024.2112$ & $ +11.9 \pm  2.4$ & MALLS \\
      $2024.2331$ & $ +14.3 \pm  6.8$ & MALLS \\
      $2024.2604$ & $ +17.0 \pm  3.3$ & MALLS \\
      $2024.2960$ & $ +15.3 \pm  3.6$ & MALLS \\
      $2024.8468$ & $  -9.1 \pm  1.6$ & GAOES-RV \\
      $2024.8495$ & $  -9.2 \pm  1.0$ & GAOES-RV \\
      $2024.9066$ & $  -0.1 \pm  5.7$ & MALLS \\
      $2024.9669$ & $  +9.0 \pm  1.5$ & GAOES-RV \\
      $2025.0246$ & $ +12.7 \pm  1.1$ & GAOES-RV \\
      $2025.1014$ & $ +14.5 \pm  1.2$ & GAOES-RV \\
      $2025.2615$ & $  +4.8 \pm  0.4$ & HDS \\
      \hline
  \end{tabular}} \label{tab:RV_g1010}
\end{table}

We assume that the RVs of the primary star follows the Keplarian
motion. We obtain its orbital elements, using a Markov Chain Mote
Carlo (MCMC) method. Its free parameters we adopt are the RV
semi-amplitude of the primary star ($\kvisi$), outer orbital period
($\pout$), outer orbital eccentricity ($\eout$), argument of
periastron of the primary star ($\argpr$), outer periastron time
($\tprimine$), and center-of-mass RV of the system ($\vcom$). Note
that the reference epoch is 2023.5 in the Julian year. We set uniform
priors on all parameters in the following ranges:
\begin{itemize}
\item $\displaystyle \left[ 0.4\left( \max {\rm RV}_i - \min {\rm
    RV}_i \right), 0.6\left( \max {\rm RV}_i - \min {\rm RV}_i \right)
  \right]$ for $\kvisi$, where ${\rm RV}_i$ is a RV we measure at each
  epoch.
\item $\displaystyle \left[ 0.8 P_{\rm Gaia}, 1.2 P_{\rm Gaia}
  \right]$ for $\pout$, where $P_{\rm Gaia}$ is an orbital period
  indicated in Gaia DR3.
\item No constraint, or $\displaystyle \left[ 0, 1 \right]$ for $\eout$.
\item No constraint for $\argpr$.
\item No constraint for $\tprimine$.
\item $\displaystyle \left[ \min {\rm RV}_i, \max {\rm RV}_i \right]$
  for $\vcom$.
\end{itemize}
The likelihood function can be expressed as
\begin{align}
  \log L = -\frac{1}{2} \sum_i \frac{({\rm RV}_{{\rm pred},i}-{\rm
      RV}_{i})^2}{\sigma_{{\rm RV},i}^2},
\end{align}
where ${\rm RV}_{{\rm pred},i}$ is the predicted RV at the same epoch
as ${\rm RV}_{i}$, and $\sigma_{{\rm RV},i}$ is the error associated
with the ${\rm RV}_{i}$ measurement. The actual values of ${\rm
  RV}_{i}$ and $\sigma_{{\rm RV},i}$ are shown in Table
\ref{tab:RV_g1010}. We implement the above MCMC method with the help
of emcee \citep{2013PASP..125..306F}.

\begin{figure*}
 \begin{center}
   \includegraphics[width=16cm]{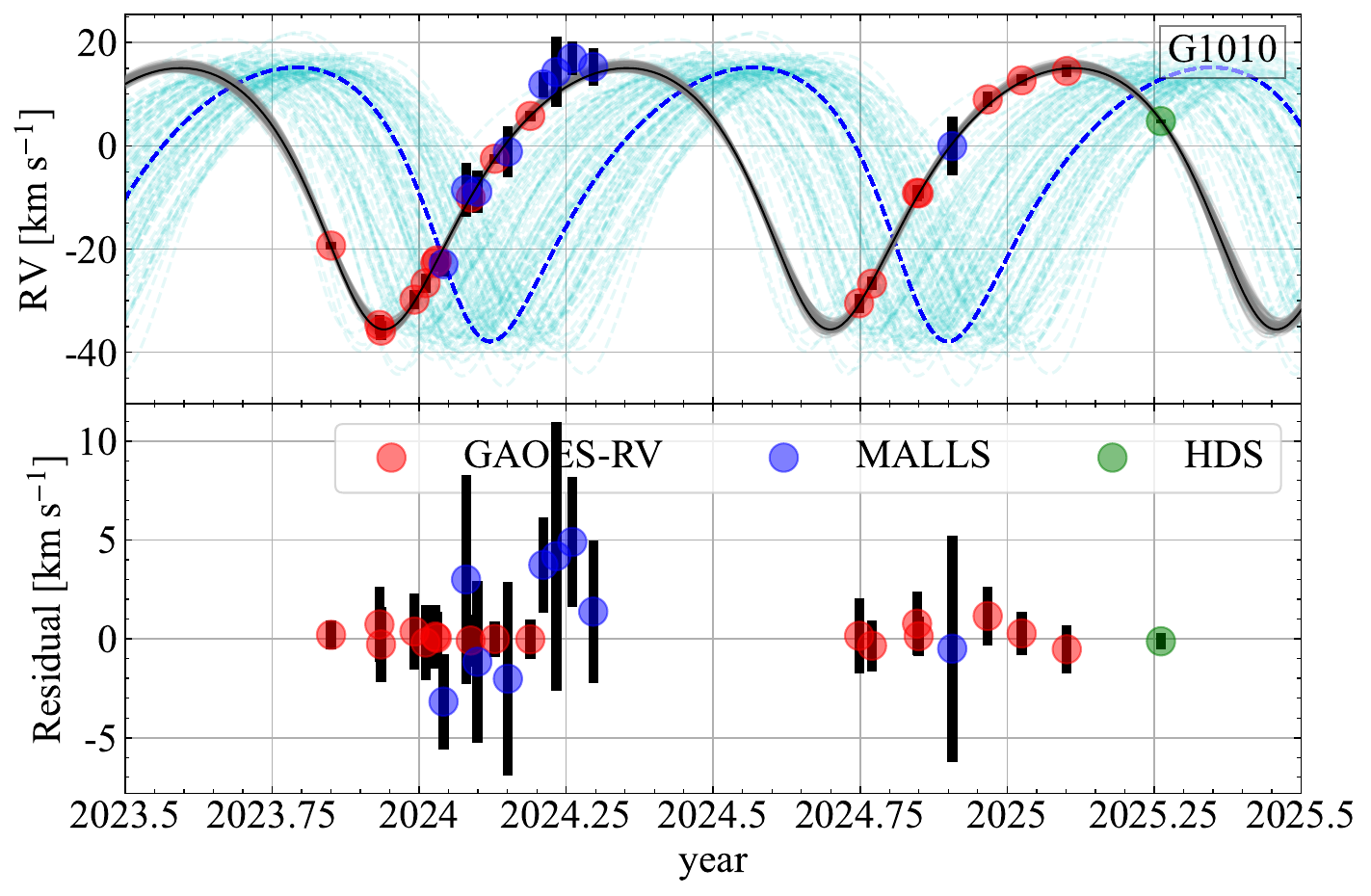}
 \end{center}
 \caption{RV variation of the primary star of \tprim. In the top
   panel, our best-fitting model is indicated by the solid thick
   curve. The $1\sigma$-confidence models are drawn by the solid thin
   curves. The dashed thick and thin curves show the best-fitting and
   $1\sigma$-confidence models provided by Gaia DR3. In the bottom
   panel, RV residuals from our best-fitting model are shown.}
\label{fig:RVs}
\end{figure*}

In Figure \ref{fig:RVs}, we draw our best-fitting and
$1\sigma$-confidence models for the RVs of the primary star in
\tprim. We also summarize their orbital elements in Table
\ref{tab:Orbitalparameters} (see also Table \ref{tab:Summary}). For
comparison, we also show RV variations (Figure \ref{fig:RVs}) and
orbital elements (Table \ref{tab:Orbitalparameters}) derived from Gaia
DR3. We show the outer periastron time, $\tprigaia$ and $\tprimine$,
setting the reference epoch to $2016$ and $2023.5$ in the Julian year,
respectively. We find that the outer orbital elements of \tprim are
consistent between Gaia DR3's and our models within $1\sigma$
confidence. Although Gaia DR3's and our RV variations look different
in Figure \ref{fig:RVs}, it is only a phase difference. Since the
orbital elements derived from Gaia DR3 are based on observational data
during 2014 -- 2017, a small error in the periastron time increases
during 2023.5 -- 2025.5 as seen in $\tprigaia$ and $\tprimine$ of
Table \ref{tab:Orbitalparameters}.

For reference, we indicate spectroscopic mass functions
($\fmas$) in Table \ref{tab:Orbitalparameters}. The spectroscopic mass
functions can be given by 
\begin{align}
  \fmas &= \frac{(\masss+\masst)^3}{(\massp+\masss+\masst)^2} \sin^3
  \iout \nonumber \\
  &= 1\;\msun \; \left( \frac{\kvisi}{\kbase\;{\rm km/s}} \right)^3
  \left( \frac{\pout}{1\;{\rm year}} \right) \left( 1-\eout^2
  \right)^{3/2}, \label{eq:Massfunction}
\end{align}
where $\iout$ is the inclination angle between our line of sight, and
the direction of the orbital angular momentum of a binary star. We
find that our spectroscopic mass function
($0.431_{-0.031}^{+0.033}\;\msun$) is consistent with that derived
from Gaia DR3 ($0.485_{-0.070}^{+0.096}\;\msun$) within $1\sigma$.

\begin{longtable}{lcccccccc}
  \caption{Comparison between orbital parameters of this work and Gaia
    DR3 for \tprim.}\label{tab:Orbitalparameters}
  \hline\noalign{\vskip3pt}
   & $\kvisi$ & $\pout$ & $\eout$ & $\argpr$ & $\tprigaia$
  & $\tprimine$ & $\vcom$ & $\fmas$ \\ [2pt]
  & [km/s]   & [day]     &        & [deg]    & [day]
         & [day]            & [km/s]  & [$\msun$] \\ [2pt]
  \hline\noalign{\vskip3pt} 
  \endfirsthead      
  \hline\noalign{\vskip3pt} 
  \hline\noalign{\vskip3pt} 
  \endhead
  \hline\noalign{\vskip3pt} 
  \endfoot
  \hline\noalign{\vskip3pt} 
  \endlastfoot
  \rule{0pt}{12pt}
  This work & $25.36_{-0.66}^{+0.69}$ & $277.2_{-1.3}^{+1.6}$
  & $0.230_{-0.018}^{+0.019}$ & $166.2_{-6.2}^{+6.0}$ &
  $119.7_{-18.9}^{+10.1}$ & $-122.0_{-2.9}^{+3.2}$ &
  $-4.25_{-0.33}^{+0.34}$ & $0.431_{-0.031}^{+0.033}$ \\
  \rule{0pt}{12pt}
  Gaia DR3 & $26.43_{-1.44}^{+1.87}$ & $284.4_{-2.9}^{+2.6}$ &
  $0.275_{-0.054}^{+0.051}$ & $157.7_{-20.0}^{+20.9}$ &
  $115.3_{-14.0}^{+14.2}$ & $-64.7_{-32.9}^{+36.6}$ &
  $-4.73_{-0.91}^{+1.02}$ & $0.485_{-0.070}^{+0.096}$ \\
\end{longtable}

\section{Identification of the Inner Binary with HDS Spectroscopy}
\label{sec:sb3}

To assess the possibility that the faint companion is itself an
unresolved inner binary, we analyze the six reddest HDS orders, where
a putative inner binary would contribute a larger fractional flux
relative to the primary star (Section~\ref{ssec:sb3_spec}). An initial
one-star (SB1) model leaves two additional sets of absorption lines
visible in the cross-correlation function (CCF), prompting us to adopt
a three-star (SB3) model that successfully reproduces all three sets
of lines. Using the atmospheric parameters and flux ratios inferred
from the SB3 spectral fit, we perform an isochrone analysis to
estimate the masses of the three components and the systemic velocity
of the triple system (Section~\ref{ssec:sb3_iso}). These
spectroscopically inferred properties are consistent with those
independently derived from the outer orbit
(Section~\ref{sec:Outer_orbit}). Together, these results strongly
support the interpretation that the system is a hierarchical triple
consisting of the primary star and an inner binary. The code used for
the analysis presented in this section and the HDS spectrum is
available on
GitHub.\footnote{\url{https://github.com/kemasuda/g1010/tree/main}}

\subsection{Modeling of the HDS Spectrum}\label{ssec:sb3_spec}

\begin{figure*}
 \begin{center}
   \includegraphics[width=16cm]{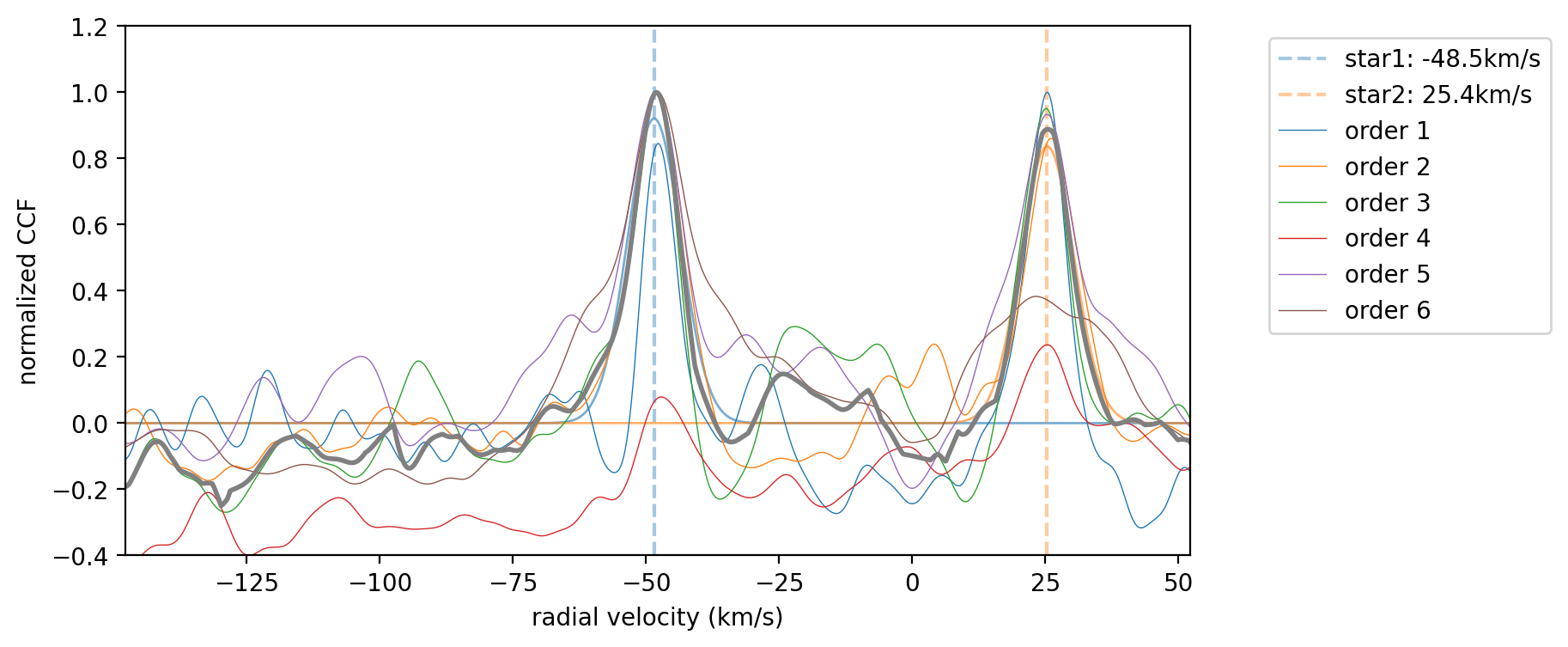}
 \end{center}
 \caption{Normalized CCFs between the solar-like template and the
   residual of the SB1 fit described in Section
   \ref{ssec:sb3_spec}. Different colors represent different
   orders. The thick gray line shows the median of the six CCFs.}
\label{fig:ccf_residual}
\end{figure*}

We fit the HDS spectrum with synthetic models using {\tt
  jaxspec}.\footnote{\url{https://github.com/kemasuda/jaxspec}} The
code interpolates the synthetic grid, Doppler-shifts it according to
the RV $v$, and convolves it with a line-broadening kernel that
includes the projected rotational velocity $v\sin i$, macroturbulence
$\zeta$, quadratic limb-darkening coefficients ($q_a$, $q_b$)
following \citet{2013MNRAS.435.2152K}, and a Gaussian instrumental
profile corresponding to $R=90,000$.  Further details of the
implementation can be found in \citet{2024ApJ...977..151T}.  Unlike in
that paper, here we employ the synthetic grid of
\citet{2005A&A...443..735C}, computed as a function of effective
temperature $T_\mathrm{eff}$, surface gravity $\log g$, metallicity
$\mathrm{[Fe/H]}$, and $\alpha$-enhancement $\mathrm{[\alpha/Fe]}$.
This grid was generated with the PFANT LTE spectral-synthesis code
using ATLAS9 model atmospheres \citep{2003IAUS..210P.A20C} and
adopting the solar abundances of \citet{1998SSRv...85..161G}.

We first perform a SB1 fit to the six HDS orders following the
procedure described in \citet{2024ApJ...977..151T}, finding an RV of
$\sim 4.7\,\mathrm{km/s}$, consistent with the outer-orbit analysis
(Section~\ref{sec:Outer_orbit}).  We then compute the residual
spectrum and cross-correlated it with a solar-like template.  The
resulting CCF exhibits two additional peaks at $\sim
-48.5\,\mathrm{km/s}$ and $\sim 25.4\,\mathrm{km/s}$, both with nearly
equal amplitudes (Figure~\ref{fig:ccf_residual}).  These two sets of
lines are also visually apparent in the middle panels of
Figure~\ref{fig:specfit_sb3}, where the primary star model from the
SB3 fit (see below) is subtracted from the data. This pattern strongly
suggests the presence of a relatively close, nearly equal-mass binary,
which we refer to as secondary and tertiary stars in the following.

Motivated by this, we refit the same spectral orders with a SB3 model,
in which the normalized flux model $F_\mathrm{model}$ is represented
as the sum of three Doppler-shifted and broadened synthetic spectra:
\begin{align}
\notag F_\mathrm{model}(\lambda) = &\left[(1 - f_2 - f_3)F_1(\lambda)
  + f_2 F_2(\lambda) + f_3 F_3(\lambda)\right] \\ &\times \left(c_a +
c_b\,\frac{\lambda - \lambda_{\rm mean}}{\lambda_{\rm max} -
  \lambda_{\rm min}}\right),
\end{align}
where $\lambda_{\rm mean}$, $\lambda_{\rm max}$, and $\lambda_{\rm
  min}$ are the mean, maximum, and minimum wavelengths within each
order. The parameters $c_a$ and $c_b$ account for the normalization
and slope in the data.  The flux ratios $f_2$ and $f_3$ were assumed
to be constant with wavelength over the limited range of the orders
analyzed.  The model was computed for each order labeled by $k$, for
which we adopt the following Gaussian-process likelihood:
\begin{equation}
\ln{\mathcal{L}^{(k)}} = -\frac{1}{2}
(F_{\rm obs}^{(k)} - F_{\rm model}^{(k)})^T
\Sigma^{-1}
(F_{\rm obs}^{(k)} - F_{\rm model}^{(k)})
-\frac{1}{2} \ln{|2\pi\Sigma|},
\end{equation}
where the covariance matrix between pixels $i$ and $j$ is
\begin{align}
\Sigma_{ij}
&= (\sigma_i^2 + s^2)\,\delta_{ij} \\
&\quad + \rho^2
\left(1 + \frac{\sqrt{3}|\lambda_i - \lambda_j|}{l}\right)
\exp\left(-\frac{\sqrt{3}|\lambda_i - \lambda_j|}{l}\right), \nonumber
\end{align}
with $\sigma_i$ fixed at $5 \times 10^{-3}$ based on the typical SNR,
$s$ accounting for any excess white noise, $\rho$ the covariance
amplitude, and $l$ the correlation length.  The total log-likelihood
is $\ln\mathcal{L} = \sum_k \ln\mathcal{L}^{(k)}$.  We first minimize
$-\ln\mathcal{L}$ using stochastic variational inference (SVI), and
then sample from the posterior using the No-U-Turn Sampler
\citep{2011arXiv1111.4246H} as implemented in {\tt NumPyro}
\citep{bingham2018pyro, phan2019composable} adopting the priors as
listed in Table~\ref{tab:specfit_sb3}.  We run four independent
chains, obtained 500--5000 effective samples for all parameters, and
checked adequate mixing using standard convergence diagnostics, finding
split Gelman--Rubin statistics $\hat{R}<1.01$ \citep{BB13945229}.

The resulting constraints are summarized in
Table~\ref{tab:specfit_sb3}.  The superscripts $(1)$-$(6)$ indicate
parameters independently fitted for each order.  The metallicity and
$\alpha$-enhancement are assumed to be common among all three stars,
and the noise parameters were assumed to be common among the six
orders.  The mean model is compared with the data in
Figure~\ref{fig:specfit_sb3}.  The middle panel in each set shows the
residuals relative to the primary star alone, and the bottom
panel shows the residuals relative to the full SB3 model.  The two
sets of residual features producing the CCF peaks
(Figure~\ref{fig:ccf_residual}) match the spectra of two $\sim
4500\,\mathrm{K}$ stars, each contributing $\lesssim 10\%$ of the
total flux.

\begin{figure*}
 \begin{center}
   \includegraphics[width=0.49\textwidth]{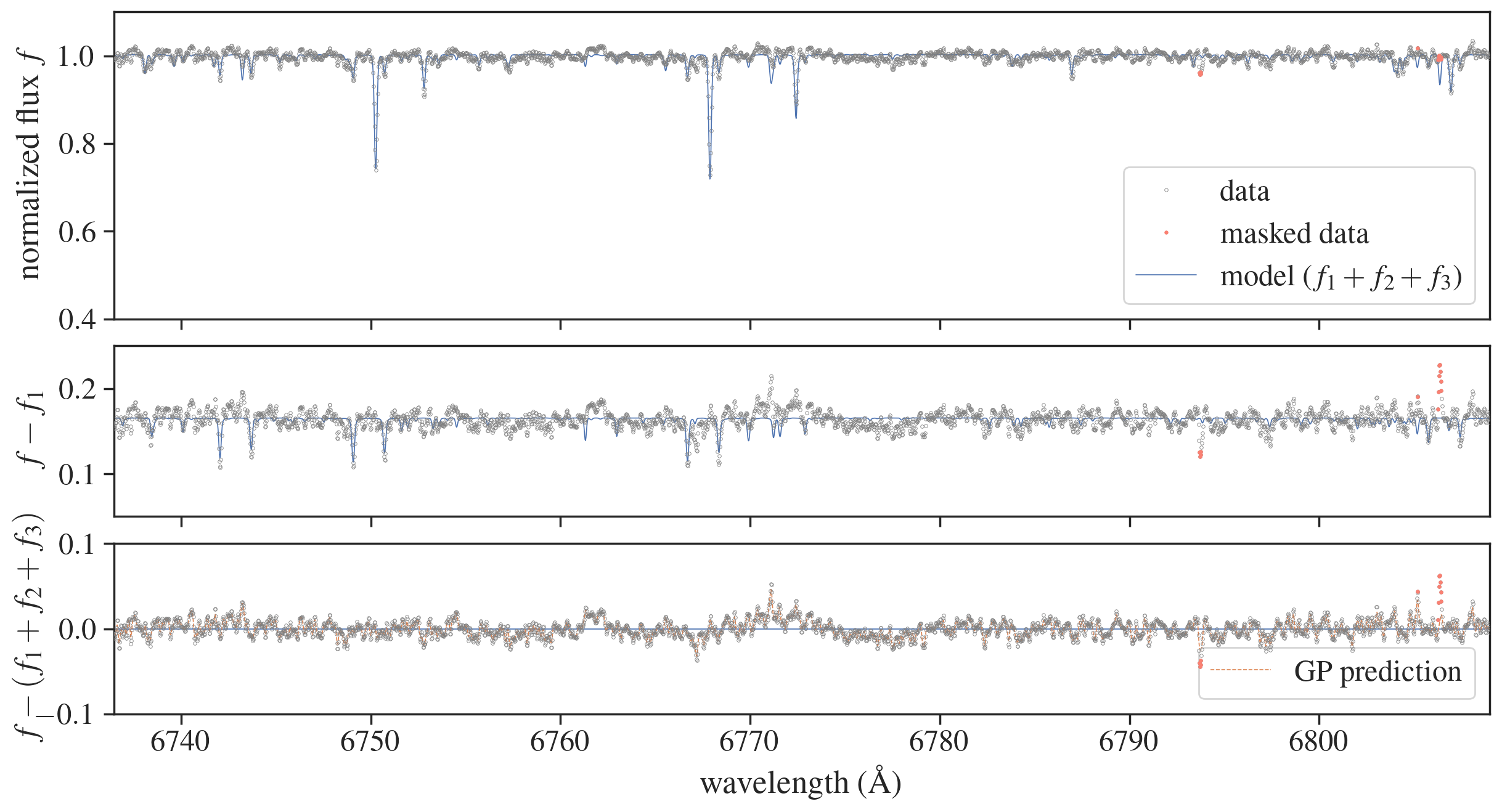}
   \includegraphics[width=0.49\textwidth]{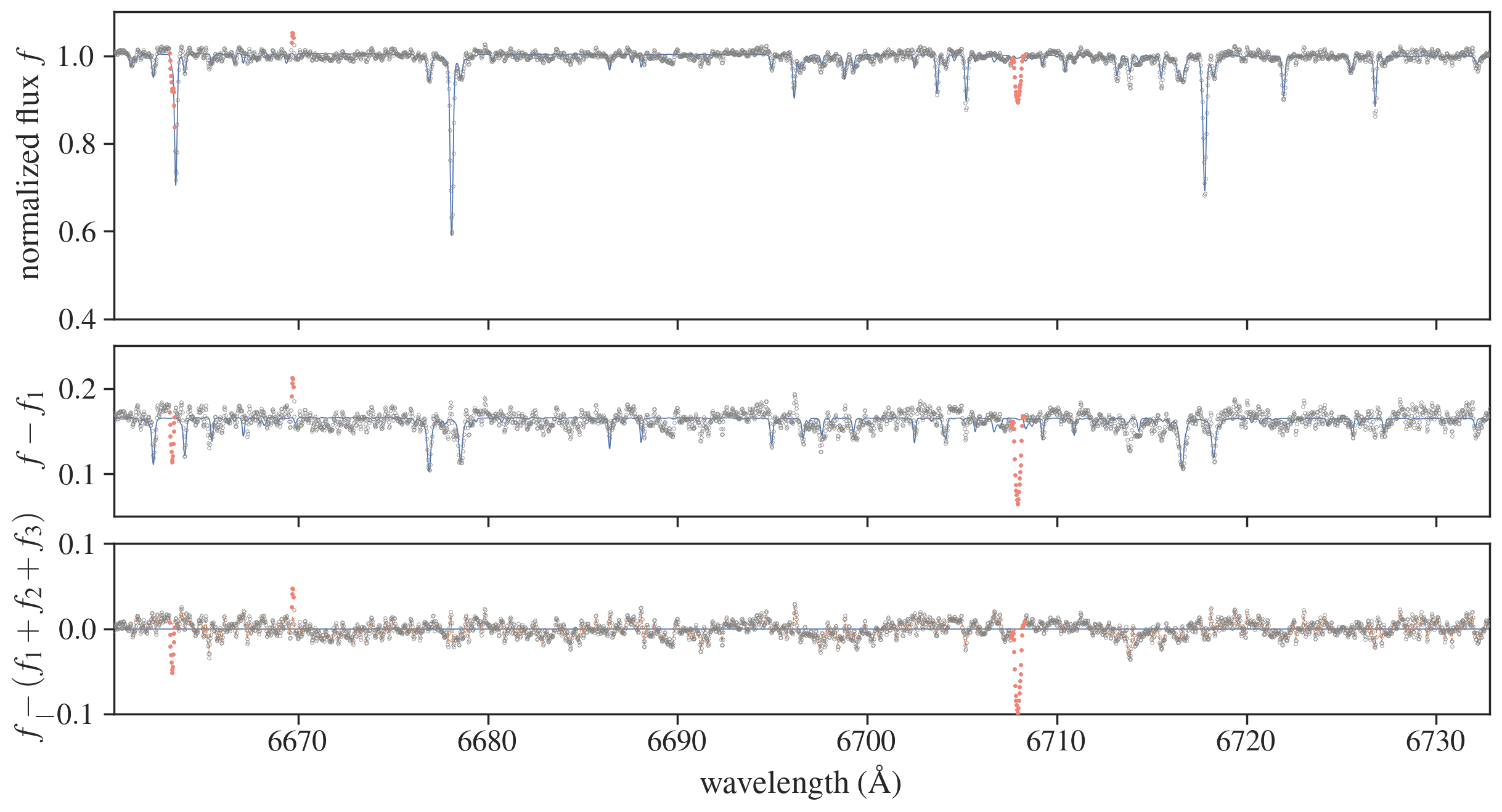}
   \includegraphics[width=0.49\textwidth]{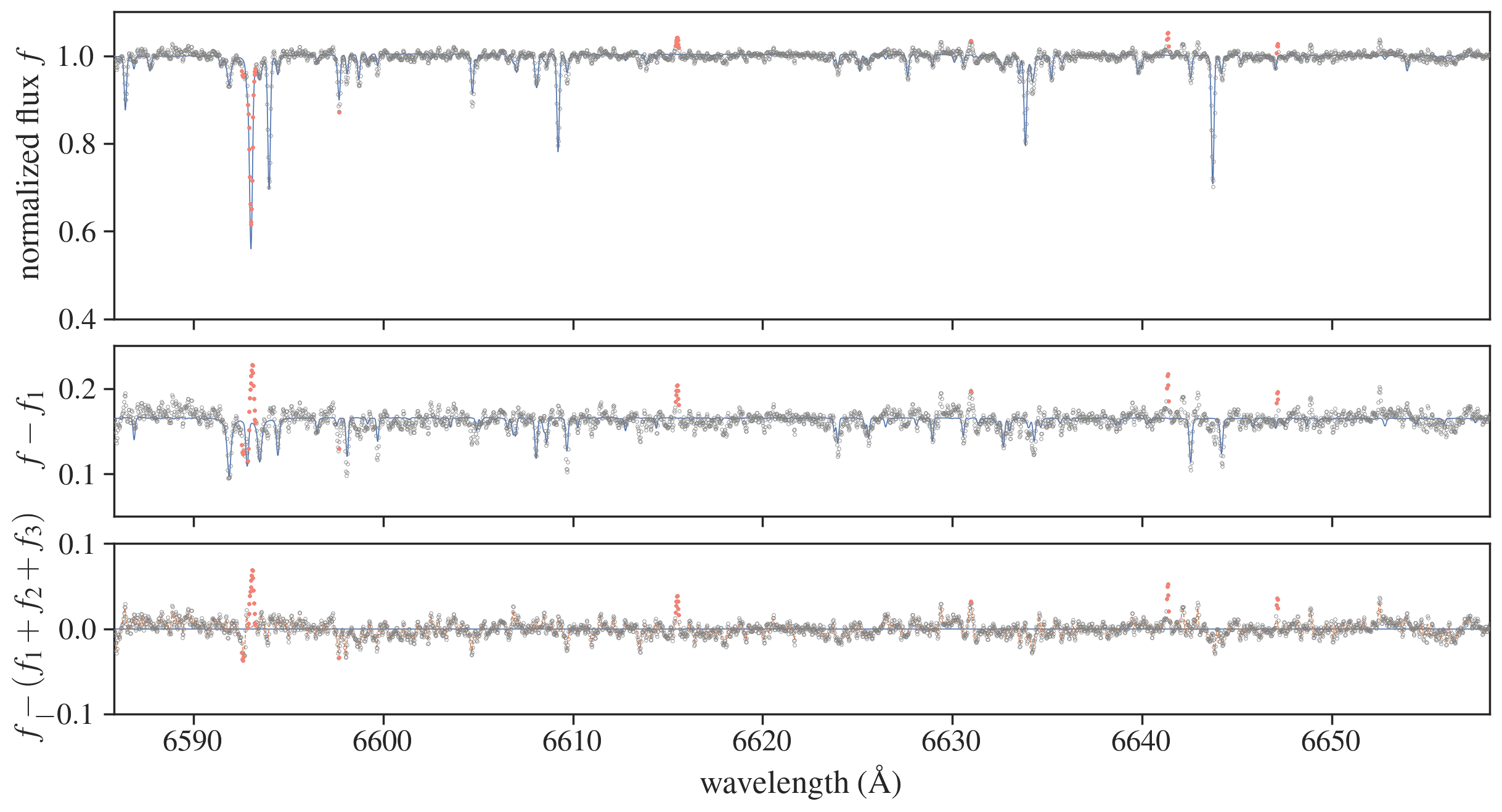}
   \includegraphics[width=0.49\textwidth]{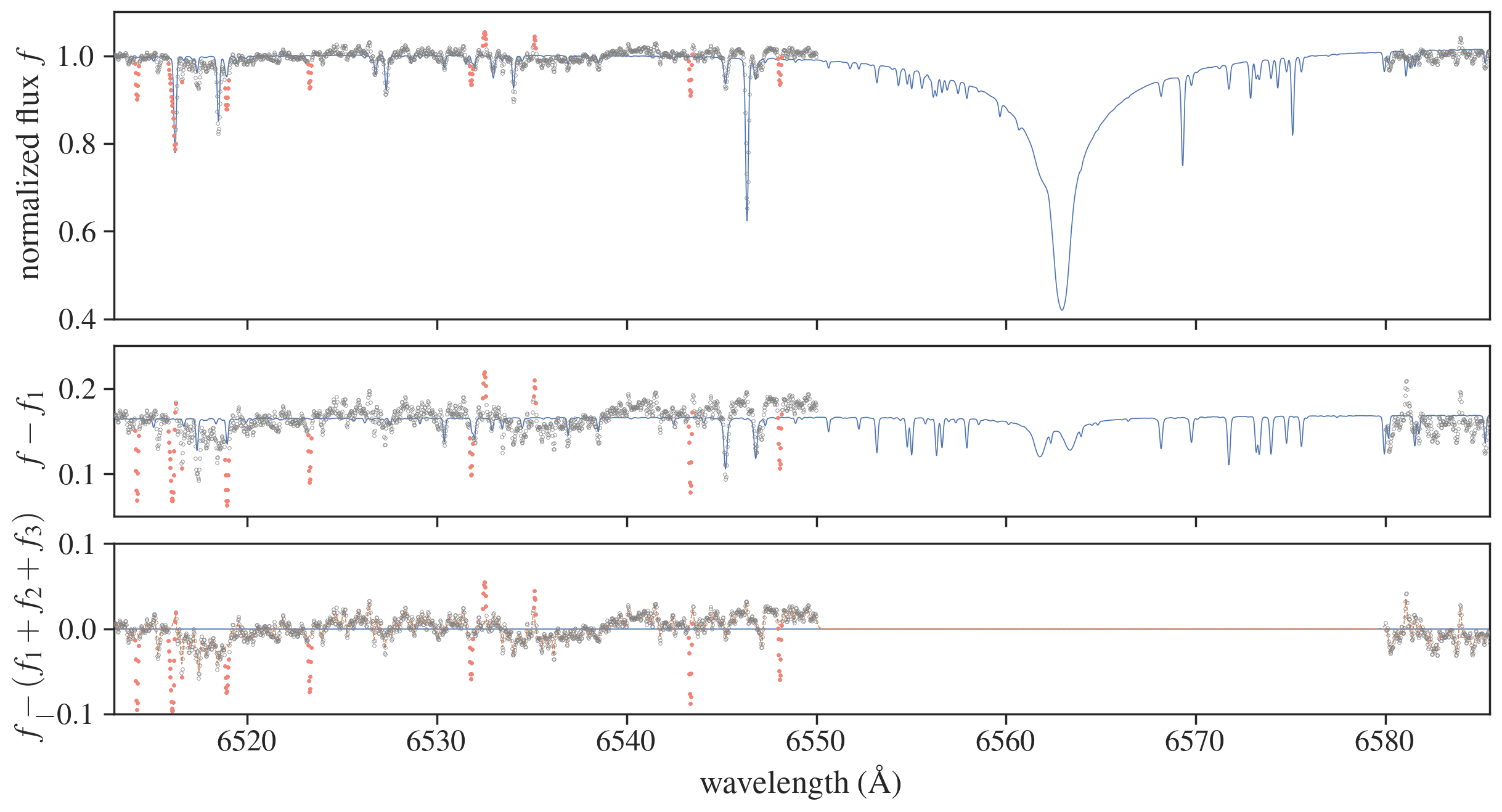}
   \includegraphics[width=0.49\textwidth]{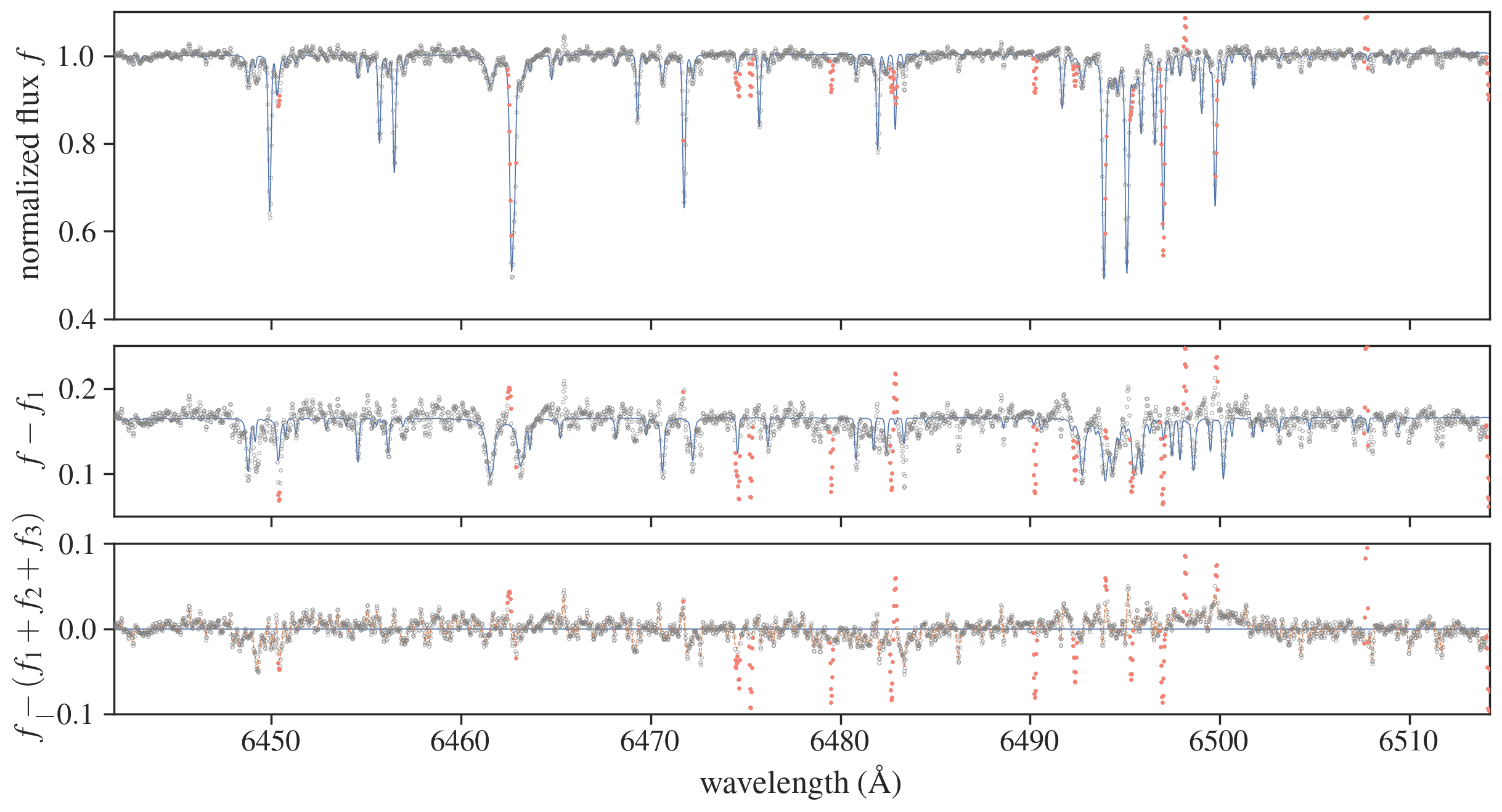}
   \includegraphics[width=0.49\textwidth]{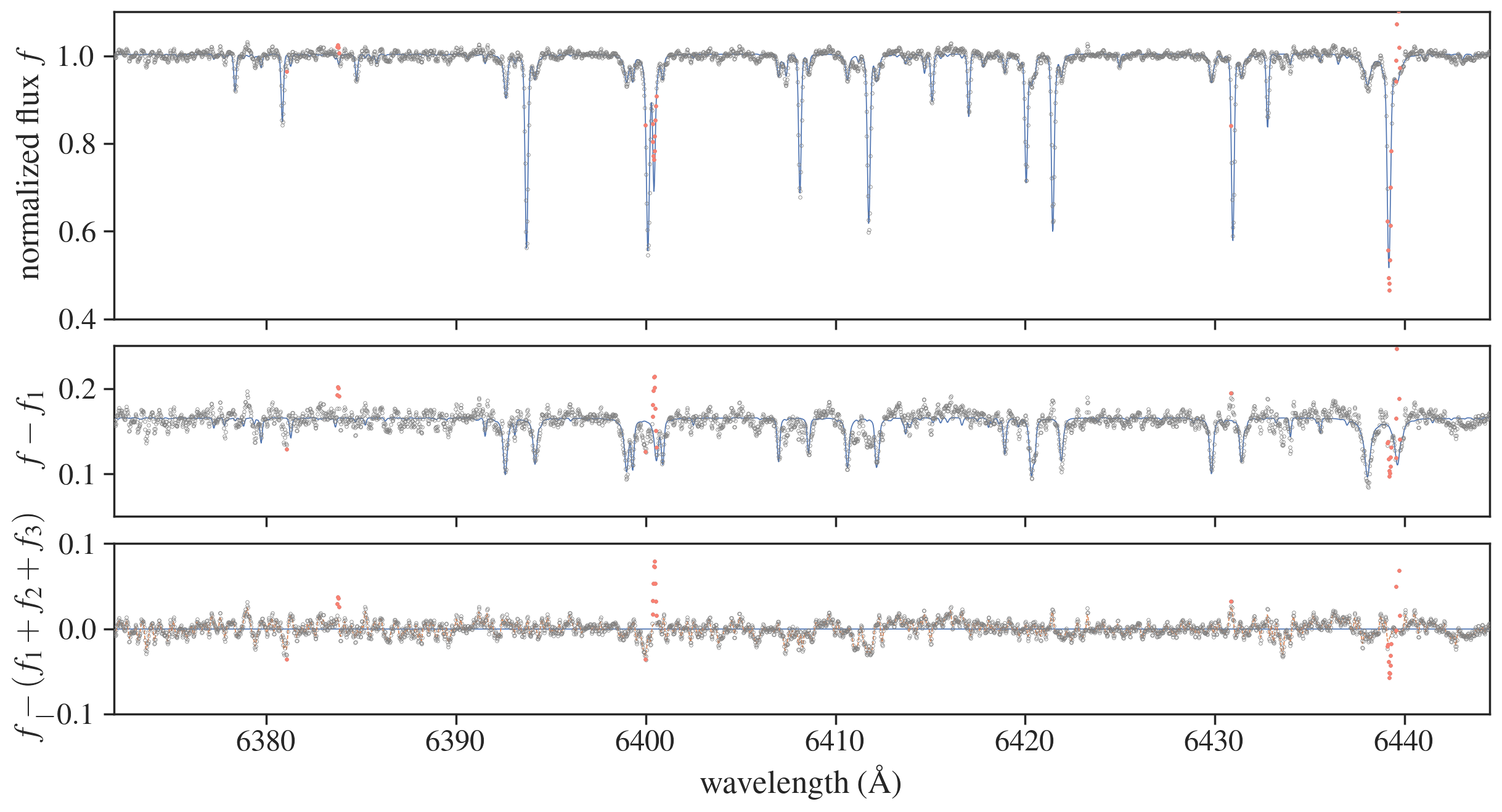}
 \end{center}
 \caption{The reddest six orders of the HDS spectrum (gray circles)
   and the spectral model (blue lines) from
   Section~\ref{ssec:sb3_spec}. Each three-panel set shows the result
   for one order. In each set, the top panel compares the synthesized
   SB3 spectrum with the normalized HDS flux, while the middle and
   bottom panels show the residuals obtained by subtracting the
   synthesized SB1 and SB3 spectra from the normalized HDS flux,
   respectively. In the bottom panels, the orange dotted lines
   indicate the mean prediction of the Gaussian-process component, and
   the filled red circles mark data points that are identified as
   outliers and excluded from the fit.}
\label{fig:specfit_sb3}
\end{figure*}

\subsection{Three-star Isochrone Fitting}\label{ssec:sb3_iso}

Given the constraints on the atmospheric parameters of the three
stars, as well as their flux ratios, we next perform an isochrone
fitting to the unresolved SB3 system to infer the physical parameters
of the components.  We use the {\tt jaxstar} code described in
\citet{2022ApJ...937...94M},\footnote{\url{https://github.com/kemasuda/jaxstar}}
which interpolates the MIST isochrones \citep{Paxton11,
  2013ApJS..208....4P, 2015ApJS..220...15P, 2016ApJ...823..102C,
  2016ApJS..222....8D} to compute the relevant stellar properties as a
function of the initial bulk metallicity $\fehi$, age $\tage$, and the
equivalent evolutionary phase eep \citep{2016ApJS..222....8D}.  The
latter essentially corresponds to the stellar mass $m$ given $\fehi$
and $\tage$.

The model parameters are the eep values of the three stars; a common
$\tage$ and $\fehi$ for the system; and the distance $d$.  The fitted
observables are the extinction-corrected $K_s$ magnitude (using
$A(K_s)=0.3026$ and $\ebv=0.107$), the Gaia DR3 parallax $\varpi$, the
effective temperatures $T_\mathrm{eff}$ of the three stars, a common
photospheric $\mathrm{[Fe/H]}$, and the flux ratios $f_2$ and $f_3$ in
the HDS wavelength range spanning $638$--$681$ nm. Note that we obtain
$\ebv$, adopting the extinction factor $\Rv = 3.1$
\citep{1989ApJ...345..245C}, and the V-band extinction $\Av$ from the
3D dust map of \citet{2022A&A...661A.147L} and
\citet{2022A&A...664A.174V} through
G-TOMO\footnote{https://explore-platform.eu/sda/g-tomo}. The model
$K_{\rm s}$ magnitude is taken to be the sum of the contributions from
the three stars; the common $\mathrm{[Fe/H]}$ is modeled as the mean
value of the three stars; and $f_2$ and $f_3$ are computed with {\tt
  jaxspec} and the grid of \citet{2005A&A...443..735C} as the ratio of
the mean fluxes in the above range multiplied by the squared radius
ratio.  The likelihood for the observables is assumed to be an
independent Gaussian for each quantity as described in
\citet{2022ApJ...937...94M}, where the uncertainties for
$T_\mathrm{eff}$ and $\mathrm{[Fe/H]}$ are inflated to
$100\,\mathrm{K}$ and 0.1~dex, respectively. The posterior samples are
obtained using the same sampler as in the SB3 spectral fit.  We run
four independent chains, obtain 1000--3000 effective samples, and
assessed chain mixing with $\hat{R}<1.01$ for all parameters.

The results are summarized in Table~\ref{tab:isochrone}, together with
the adopted priors. The masses of the three stars are inferred to be
about $0.85$, $0.63$, and $0.61\,\msun$.  Using the posterior samples
for these parameters, together with the radial velocities from the
spectral fit (Section~\ref{ssec:sb3_spec}), we compute $\fmassin =
(m_2+m_3)^3/(m_1+m_2+m_3)^2$ for the outer orbit and the
center-of-mass velocity
\begin{align}
  \gamma_\mathrm{iso} = \frac{\massp v_1 + \masss v_2 + \masst
    v_3}{\massp + \masss + \masst}
\end{align}
of the three-body system. The resulting $\fmassin =
0.432^{+0.016}_{-0.017}\,\msun$ (68\%) agrees with the value from the
RV orbit (Table~\ref{tab:Orbitalparameters}) for $\sin i \approx 1$,
and $\gamma_\mathrm{iso} = -4.93^{+0.18}_{-0.16}\,\mathrm{km/s}$ is
consistent with the RV-based value within the $2\sigma$ level.  We
emphasize that these values are obtained entirely from the single HDS
spectrum and are therefore essentially independent of the orbital
analysis in Section~\ref{sec:Outer_orbit}. The agreement strongly
indicates that the RV motion of the primary star is caused by the two
fainter stars identified in the HDS spectrum --- that is, the system
is a G+(K+K) hierarchical triple.

\begin{table*}[t]
  \centering
  \caption{Priors and Posterior Constraints from the Spectral Analysis}
  \label{tab:specfit_sb3}
  \begin{minipage}[t]{0.48\textwidth}
    \centering
  \begin{tabular}{lcc}
    \hline
    Parameter & Posterior & Prior \\
    \hline
    \multicolumn{3}{l}{\bf (Atmospheric Parameters)}\\
    $T_{\mathrm{eff},1}$ [K] & $5749_{-14}^{+14}$ & $\mathcal{U}(3500,7000)$ \\
    $T_{\mathrm{eff},2}$ [K] & $4565_{-67}^{+58}$ & $\mathcal{U}(3500,7000)$ \\
    $T_{\mathrm{eff},3}$ [K] & $4575_{-85}^{+76}$ & $\mathcal{U}(3500,7000)$ \\
    $\log g_{1}$ & $4.382_{-0.043}^{+0.040}$ & $\mathcal{U}(3,5)$ \\
    $\log g_{2}$ & $4.33_{-0.14}^{+0.15}$ & $\mathcal{U}(3,5)$ \\
    $\log g_{3}$ & $4.30_{-0.21}^{+0.24}$ & $\mathcal{U}(3,5)$ \\
    $\mathrm{[Fe/H]}$ & $-0.447_{-0.010}^{+0.011}$ & $\mathcal{U}(-1,0.5)$ \\
    $\mathrm{[\alpha/Fe]}$ & $0.0050_{-0.0036}^{+0.0070}$ & $\mathcal{U}(0,0.4)$ \\
    $(v\sin i)_{1}$ [km/s] & $2.20_{-0.13}^{+0.14}$ & $\mathcal{U}(0,15)$ \\
    $(v\sin i)_{2}$ [km/s] & $2.89_{-0.44}^{+0.37}$ & $\mathcal{U}(0,15)$ \\
    $(v\sin i)_{3}$ [km/s] & $2.37_{-0.82}^{+0.56}$ & $\mathcal{U}(0,15)$ \\
    $\zeta_{1}$ [km/s] & $3.948_{-0.022}^{+0.021}$ & $\mathcal{N}(\zeta_\mathrm{emp}(T_\mathrm{eff}),1)$ \\
    $\zeta_{2}$ [km/s] & $2.13_{-0.10}^{+0.09}$ & $\mathcal{N}(\zeta_\mathrm{emp}(T_\mathrm{eff}),1)$ \\
    $\zeta_{3}$ [km/s] & $2.14_{-0.13}^{+0.12}$ & $\mathcal{N}(\zeta_\mathrm{emp}(T_\mathrm{eff}),1)$ \\
    $q_{a,1}$ & $0.44_{-0.31}^{+0.36}$ & $\mathcal{U}(0,1)$ \\
    $q_{a,2}$ & $0.51_{-0.35}^{+0.33}$ & $\mathcal{U}(0,1)$ \\
    $q_{a,3}$ & $0.51_{-0.35}^{+0.33}$ & $\mathcal{U}(0,1)$ \\
    $q_{b,1}$ & $0.43_{-0.30}^{+0.36}$ & $ \mathcal{U}(0,1)$ \\
    $q_{b,2}$ & $0.51_{-0.36}^{+0.34}$ & $ \mathcal{U}(0,1)$ \\
    $q_{b,3}$ & $0.51_{-0.34}^{+0.33}$ & $ \mathcal{U}(0,1)$ \\
    \multicolumn{3}{l}{\bf (Flux Ratios)}\\
    $f_2$ & $0.0923_{-0.0031}^{+0.0032}$ & $\mathcal{U}(0,0.25)$ \\
    $f_3$ & $0.0727_{-0.0031}^{+0.0032}$ & $\mathcal{U}(0,0.25)$ \\
    \multicolumn{3}{l}{\bf (Radial Velocities)}\\
    RV$_{1}^{(1)}$ [km/s]& $4.833_{-0.062}^{+0.057}$ & $\mathcal{U}(-0.21,9.78)$ \\
    RV$_{1}^{(2)}$ [km/s]& $4.920_{-0.049}^{+0.046}$ & $\mathcal{U}(-0.21,9.78)$ \\
    RV$_{1}^{(3)}$ [km/s]& $4.707_{-0.048}^{+0.049}$ & $\mathcal{U}(-0.21,9.78)$ \\
    RV$_{1}^{(4)}$ [km/s]& $4.783_{-0.073}^{+0.072}$ & $\mathcal{U}(-0.21,9.78)$ \\
    RV$_{1}^{(5)}$ [km/s]& $4.735_{-0.032}^{+0.032}$ & $\mathcal{U}(-0.21,9.78)$ \\
    RV$_{1}^{(6)}$ [km/s]& $4.702_{-0.030}^{+0.029}$ & $\mathcal{U}(-0.21,9.78)$ \\
    \hline
  \end{tabular}
  \end{minipage}
  \hfill
  \begin{minipage}[t]{0.48\textwidth}
    \centering
  \begin{tabular}{lcc}
    \hline
    Parameter & Posterior & Prior \\
    \hline
    RV$_{2}^{(1)}$ [km/s]& $-47.67_{-0.23}^{+0.23}$ & $\mathcal{U}(-0.21,9.78)-53.2$ \\
    RV$_{2}^{(2)}$ [km/s]& $-48.41_{-0.29}^{+0.29}$ & $\mathcal{U}(-0.21,9.78)-53.2$ \\
    RV$_{2}^{(3)}$ [km/s]& $-47.84_{-0.21}^{+0.21}$ & $\mathcal{U}(-0.21,9.78)-53.2$ \\
    RV$_{2}^{(4)}$ [km/s]& $-47.24_{-0.32}^{+0.33}$ & $\mathcal{U}(-0.21,9.78)-53.2$ \\
    RV$_{2}^{(5)}$ [km/s]& $-48.02_{-0.16}^{+0.17}$ & $\mathcal{U}(-0.21,9.78)-53.2$ \\
    RV$_{2}^{(6)}$ [km/s]& $-47.87_{-0.22}^{+0.23}$ & $\mathcal{U}(-0.21,9.78)-53.2$ \\
    RV$_{3}^{(1)}$ [km/s]& $25.67_{-0.26}^{+0.29}$ & $\mathcal{U}(-0.21,9.78)+20.6$ \\
    RV$_{3}^{(2)}$ [km/s]& $26.17_{-0.32}^{+0.33}$ & $\mathcal{U}(-0.21,9.78)+20.6$ \\
    RV$_{3}^{(3)}$ [km/s]& $25.71_{-0.24}^{+0.24}$ & $\mathcal{U}(-0.21,9.78)+20.6$ \\
    RV$_{3}^{(4)}$ [km/s]& $25.82_{-0.37}^{+0.33}$ & $\mathcal{U}(-0.21,9.78)+20.6$ \\
    RV$_{3}^{(5)}$ [km/s]& $26.04_{-0.20}^{+0.18}$ & $\mathcal{U}(-0.21,9.78)+20.6$ \\
    RV$_{3}^{(6)}$ [km/s]& $25.79_{-0.27}^{+0.25}$ & $\mathcal{U}(-0.21,9.78)+20.6$ \\
    \multicolumn{3}{l}{\bf (Noise Parameters)}\\
    $\ln \rho$ & $-4.643_{-0.014}^{+0.014}$ & $ \mathcal{U}(-5,-0.5)$ \\
    $\ln l$ [\AA] & $-2.168_{-0.017}^{+0.018}$ & $ \mathcal{U}(-5,2)$ \\
    $\ln s$ & $-9.67_{-0.23}^{+0.35}$ & $ \mathcal{U}(-10,-3)$ \\
    \multicolumn{3}{l}{\bf (Flux Normalization and Slope)}\\
    $c_{a}^{(1)}$ & $1.00268_{-0.00060}^{+0.00064}$ & $\mathcal{U}(0.8,1.2)$ \\
    $c_{a}^{(2)}$ & $1.00280_{-0.00064}^{+0.00061}$ & $\mathcal{U}(0.8,1.2)$ \\
    $c_{a}^{(3)}$ & $1.00317_{-0.00062}^{+0.00060}$ & $\mathcal{U}(0.8,1.2)$ \\
    $c_{a}^{(4)}$ & $1.01009_{-0.00091}^{+0.00093}$ & $\mathcal{U}(0.8,1.2)$ \\
    $c_{a}^{(5)}$ & $1.00387_{-0.00062}^{+0.00064}$ & $\mathcal{U}(0.8,1.2)$ \\
    $c_{a}^{(6)}$ & $1.00414_{-0.00059}^{+0.00062}$ & $\mathcal{U}(0.8,1.2)$ \\
    $c_{b}^{(1)}$ & $0.0002_{-0.0020}^{+0.0020}$ & $\mathcal{U}(-0.2,0.2)$ \\
    $c_{b}^{(2)}$ & $-0.0066_{-0.0020}^{+0.0021}$ & $\mathcal{U}(-0.2,0.2)$ \\
    $c_{b}^{(3)}$ & $-0.0049_{-0.0020}^{+0.0020}$ & $\mathcal{U}(-0.2,0.2)$ \\
    $c_{b}^{(4)}$ & $0.0257_{-0.0029}^{+0.0028}$ & $\mathcal{U}(-0.2,0.2)$ \\
    $c_{b}^{(5)}$ & $0.0063_{-0.0021}^{+0.0021}$ & $\mathcal{U}(-0.2,0.2)$ \\
    $c_{b}^{(6)}$ & $0.0009_{-0.0021}^{+0.0021}$ & $\mathcal{U}(-0.2,0.2)$ \\
    \hline
  \end{tabular}
  \end{minipage}
  \begin{tabnote}
    Note --- Values listed here report the medians and $68\%$
    equal-tail intervals of the marginal posteriors. $\mathcal{U}
    (a,b)$ denotes the uniform probability density function between
    $a$ and $b$. $\mathcal{U} (\mu,\sigma)$ denotes the normal
    distribution with mean $\mu$ and standard deviation $\sigma$. The
    $\zeta_\mathrm{emp}(T_\mathrm{eff})$ was computed using the
    empirical relation (equation~1) in \citet{2005ApJS..159..141V}
    where we have corrected the sign of the linear term.
  \end{tabnote}
\end{table*}

\begin{table}
  \centering
  \caption{Priors and Posterior Constraints from the Isochrone Analysis}
   \label{tab:isochrone}
  \begin{tabular}{lcc}
    \hline
    Parameter & Posterior & Prior \\
    \hline
    distance [kpc] & $0.465_{-0.028}^{+0.029}$ & $\Gamma(3, 1.58)$ \\
    age $\tage$ [Gyr] & $12.2_{-1.9}^{+1.2}$ & $\mathcal{U}(0.1, 13.8)$\\
    $\mathrm{eep}_{1}$ & $420_{-12}^{+10}$ &  $\mathcal{U}(0, 500)$\\
    $\mathrm{eep}_{2}$ & $341.4_{-4.6}^{+3.0}$ &  $\mathcal{U}(0, 500)$\\
    $\mathrm{eep}_{3}$ & $338.4_{-4.7}^{+3.0}$ &  $\mathcal{U}(0, 500)$\\
    $\fehi$ & $-0.334_{-0.086}^{+0.087}$ &  $\mathcal{U}(-1, 0.5)$\\
    $\varpi$ [mas] & $2.15_{-0.13}^{+0.14}$ & * \\
    $T_{\mathrm{eff},1}$ & $5869_{-60}^{+65}$ & * \\
    $T_{\mathrm{eff},2}$ & $4585_{-71}^{+75}$ & * \\
    $T_{\mathrm{eff},3}$ & $4454_{-67}^{+70}$ & *\\
    $f_2$ & $0.0918_{-0.0030}^{+0.0029}$ &  *\\
    $f_3$ & $0.0736_{-0.0028}^{+0.0029}$ & *\\
    $\fehn_1$ & $-0.450_{-0.098}^{+0.100}$ &  \\
    $\fehn_2$ & $-0.385_{-0.089}^{+0.092}$ &  \\
    $\fehn_3$ & $-0.380_{-0.089}^{+0.091}$ &  \\
    mean $\fehn$ & $-0.405_{-0.092}^{+0.094}$ & * \\
    $K_{{\rm s},1}$ [mag] & $11.415_{-0.031}^{+0.031}$ &  \\
    $K_{{\rm s},2}$ [mag] & $12.955_{-0.053}^{+0.057}$ &  \\
    $K_{{\rm s},3}$ [mag] & $13.078_{-0.051}^{+0.056}$ &  \\
    total $K$ [mag] & $11.006_{-0.021}^{+0.021}$ &  *\\
    $m_{1}$ [$\msun$] & $0.851_{-0.025}^{+0.030}$ &  \\
    $m_{2}$ [$\msun$] & $0.628_{-0.020}^{+0.020}$ &  \\
    $m_{3}$ [$\msun$] & $0.606_{-0.020}^{+0.020}$ &  \\
    $R_{1}$ [$\rsun$] & $1.099_{-0.073}^{+0.082}$ &  \\
    $R_{2}$ [$\rsun$] & $0.625_{-0.020}^{+0.019}$ &  \\
    $R_{3}$ [$\rsun$] & $0.605_{-0.022}^{+0.019}$ &  \\
    $\fmassin$ [$\msun$] & $0.432^{+0.016}_{-0.017}$ &\\
    $\gamma_\mathrm{iso}$ [km/s] & $-4.93^{+0.18}_{-0.16}$ &\\
    \hline
  \end{tabular}
    \begin{tabnote}
    Note --- Values listed here report the medians and $68\%$ equal-tail intervals of the marginal posteriors. 
    Parameters with * in the prior column are the fitted observables; see Section~\ref{ssec:sb3_iso}.
  \end{tabnote}
\end{table}

\begin{figure}
 \begin{center}
   \includegraphics[width=8cm]{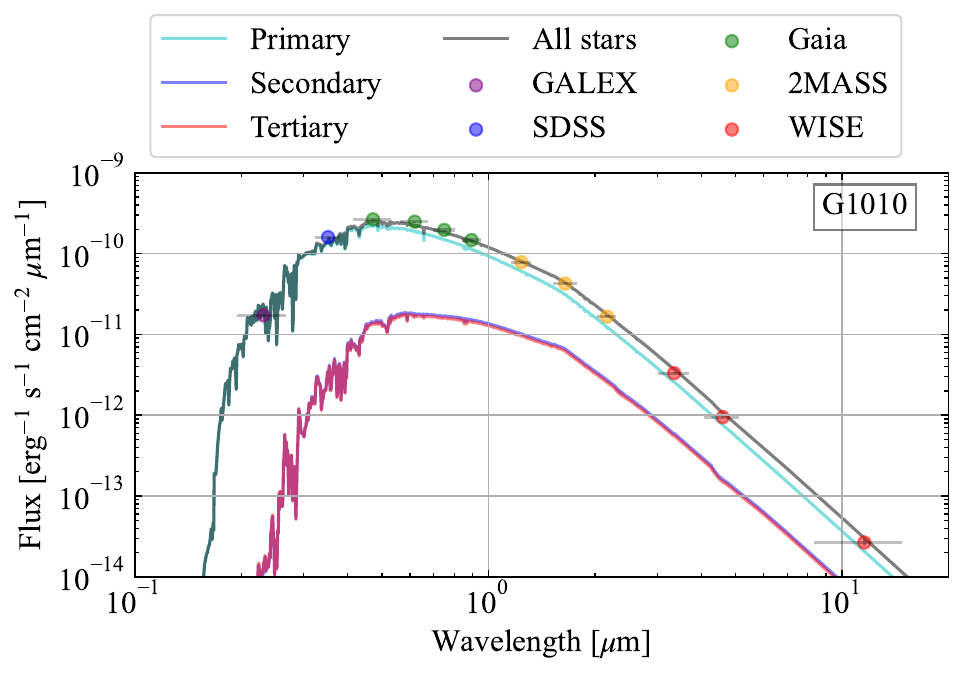}
 \end{center}
\caption{SED of \tprim. We construct these SEDs, retrieving WISE
  W$_1$W$_2$W$_3$, 2MASS JHK, u SDSS, and GALEX NUV photometry from
  VizieR, and calculating griz SDSS photometry, filtering Gaia XP
  spectra with pyphot. We synthesize several spectra by means of the
  pyphot code.}
\label{fig:SED}
\end{figure}

We synthesize the spectrum of \tprim using stellar parameters obtained
from the isochrone fitting, and compare the spectra with the
broad-band spectral energy distributions (SED) of \tprim. For the
spectral synthesis, we use the pystellibs
code\footnote{https://github.com/mfouesneau/pystellibs} with the BaSeL
library (\cite{1997A&AS..125..229L}; \yearcite{1998A&AS..130...65L}),
adopting parameters obtained from the three-star isochrone fitting. We
use the interstellar reddening $\ebv=0.107$ as described above. We
construct the broad-band SED of \tprim as follows. We adopt WISE
W$_1$W$_2$W$_3$ photometry \citep{2010AJ....140.1868W}, 2MASS JHK
photometry \citep{2006AJ....131.1163S}, u Sloan Digital Sky Survey
(SDSS) photometry \citep{2008ApJ...674.1217P} and GALEX NUV photometry
\citep{2005ApJ...619L...1M}, retrieving these data from VizieR
\citep{vizier2000}. We synthesize griz SDSS photometry
\citep{2008ApJ...674.1217P}, filtering Gaia XP spectra
\citep{2023A&A...674A..33G} with pyphot, a tool for computing
photometry from spectra \citep{zenodopyphot}. As seen Figure
\ref{fig:SED}, the synthesized spectrum is consistent with the the
broad-band SED. The residuals of the photometry lie within about $0.1$
mag.

\section{Further Evidence for the Inner Binary from TESS}
\label{sec:Inner_orbit}

\begin{figure}
 \begin{center}
    \includegraphics[width=8cm]{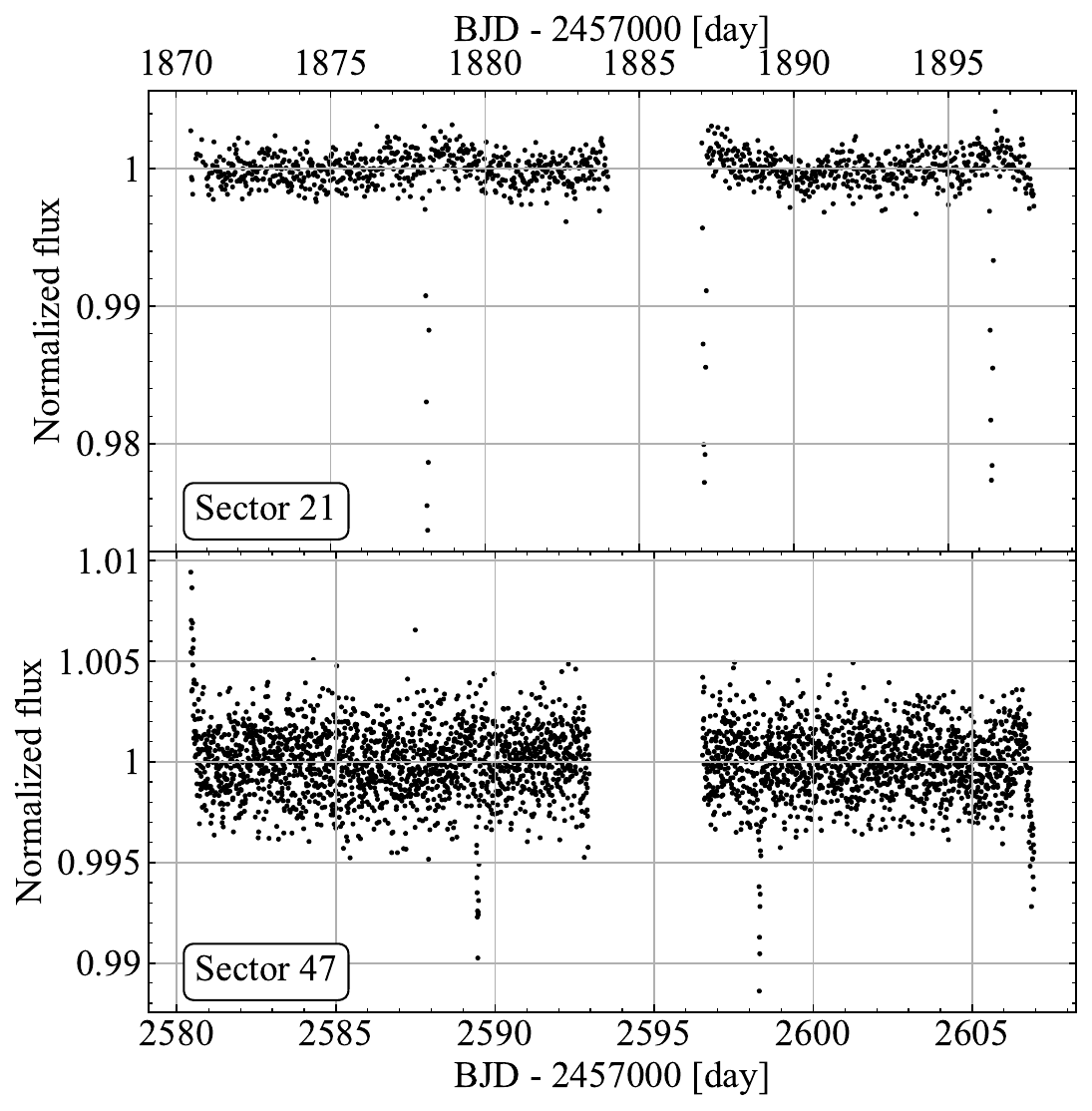}
 \end{center}
 \caption{Normalized light curves for \tprim (TIC 21502513) from TESS
   sectors 21 and 47. Nominal values are plotted.}
\label{fig:lightcurve_g1010}
\end{figure}

\begin{figure*}
 \begin{center}
    \includegraphics[width=16cm]{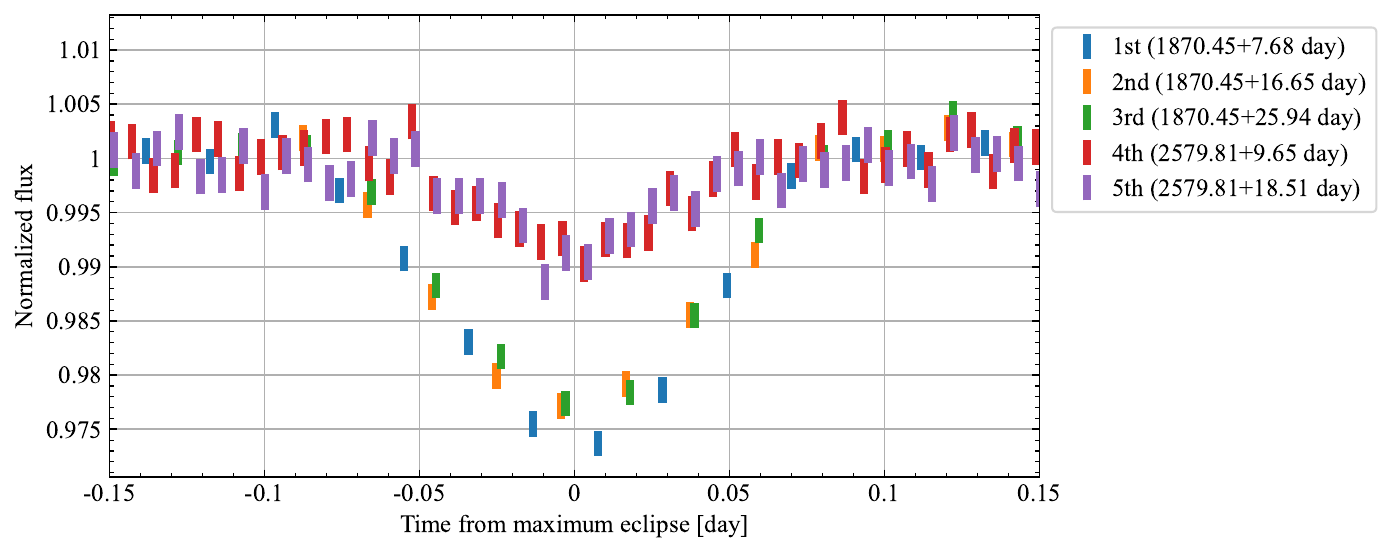}
 \end{center}
 \caption{Normalized light curves for \tprim (TIC 21502513) from TESS
   sectors 21 and 47. Sector 21 includes the 1st, 2nd, and 3rd
   eclipses, while sector 47 covers the 4th and 5th. For each event,
   the time axis is shifted to center the maximum eclipse at zero. The
   specific shift amounts, expressed in ${\rm BJD}-2457000$ [days],
   are detailed in the legend. Error bars are plotted.}
\label{fig:eclipse_g1010}
\end{figure*}

\tprim corresponds to TIC 21502513 in the TESS mission
\citep{2015JATIS...1a4003R} and is observed in sectors 21 and
47. Figures \ref{fig:lightcurve_g1010} and \ref{fig:eclipse_g1010}
displays the full light curve and the zoomed-in views of the eclipses
for this source, respectively. We identify three eclipses in sector 21
and two in sector 47. The characteristics of these eclipse events are
as follows: the first three eclipses (sector 21) exhibit similar
durations and depths, as do the final two (sector 47). However, the
eclipses in sector 21 are notably longer and deeper than those in
sector 47. The measured time intervals between successive events --
the 1st and 2nd, 2nd and 3rd, and 4th and 5th eclipses -- are
approximately 9.0, 9.3, and 8.9 days, respectively.

We infer that the three eclipses in sector 21 correspond to the
primary and secondary eclipses of the inner binary of
\tprim. We also suggest that these eclipses are diluted by
the flux of the primary star. This interpretation is supported by the
following three points, which are consistent with our findings in
Section \ref{sec:sb3}. First, the three eclipses exhibit similar
durations and depths. This observation aligns with our isochrone
fitting results (Section \ref{ssec:sb3_iso}), which indicate that the
inner binary consists of two stars with nearly identical masses and
radii.

Second, the time interval between the 1st and 3rd eclipses ($\sim
18.3$ days) is expected to correspond to the orbital period of the
inner binary. Based on this, the average relative velocity of the
inner binary components can be estimated as:
\begin{align}
  v \sim 87 \left( \frac{m_2+m_3}{1.24\;\msun} \right)^{1/3} \left(
  \frac{\pin}{18.3\;{\rm day}} \right)^{-1/3} \; [{\rm km/s}],
\end{align}
where $\pin$ denotes the inner orbital period. This value is
consistent with the relative velocity of $\sim 73.9$ km/s derived from
our SB3 spectral fit (Section \ref{ssec:sb3_spec}). The fact that the
2nd eclipse does not occur exactly halfway between the 1st and 3rd
eclipses suggests that the inner binary possesses a small orbital
eccentricity ($\sim 0.01$).

Third, the observed durations of these eclipses are approximately
$0.1$ days (see Figure \ref{fig:eclipse_g1010}). For comparison, we
can estimate the expected durations using the relative velocity
derived above and the radii of the secondary and tertiary stars
obtained from the isochrone fitting in Section \ref{ssec:sb3_iso}. The
duration $D$ is given by:
\begin{align}
  D \sim 0.11 \left( \frac{R}{0.6\;\rsun} \right) \left(
  \frac{v}{87\;{\rm km/s}} \right)^{-1} \; [{\rm day}],
\end{align}
where $R$ denotes the radius of the secondary or tertiary star. This
calculated duration is in excellent agreement with the observations
shown in Figure \ref{fig:eclipse_g1010}.

The time interval between the 4th and 5th eclipses (8.9 days) deviates
significantly from those of the 1st and 2nd (9.0 days) and the 2nd and
3rd (9.3 days) eclipses. Furthermore, the eclipse depths also differ
between sectors 21 and 47. We speculate that these discrepancies can
be attributed to three-body effects: specifically, apsidal precession
may account for the varying time intervals, while nodal precession
explains the changes in eclipse depth. The timescale for the inner
orbit of \tprim to precess by $180^\circ$ can be estimated as follows:
\begin{align}
  t_{\rm 3body} &\sim 6.7 \times 10^3
  \left( \frac{\masss+\masst}{1.24\;\msun} \right)
  \left( \frac{\massp}{0.85\;\msun} \right)^{-1}\nonumber \\
  &\times \left( \frac{a_{\rm in}}{0.147\;{\rm au}} \right)^{-3}
  \left( \frac{a_{\rm out}}{1.06\;{\rm au}} \right)^3
  \left( \frac{\pin}{18.3\;{\rm day}} \right)
  \;[{\rm day}], \label{eq:t3body}
\end{align}
assuming a circular inner orbit and coplanarity between the inner and
outer orbits (i.e., mutual inclination of zero). Given that the time
interval between sectors 21 and 47 is 710 days -- a value comparable
to $t_{\rm 3body}$ -- it is plausible that the distinct eclipse
features observed across these sectors are driven by three-body
dynamical effects.

\section{Comparison with previous discoveries}
\label{sec:Comparison}

We compare our triple system \tprim with previous
surveys. \citet{2018ApJS..235....6T} have published the Multiple Star
Catalog that includes not only compact triple star systems but also
resolved triple star systems. However, \tprim is not in the catalog.
Many studies have searched for triple star systems via eclipse time
variations from Kepler binaries (\cite{2011MNRAS.417L..31S};
\cite{2012AJ....143..137G}; \yearcite{2015AJ....150..178G};
\cite{2013ApJ...768...33R}; \cite{2013MNRAS.428.1656B}
\yearcite{2015MNRAS.448..946B}; \yearcite{2016MNRAS.455.4136B};
\yearcite{2025A&A...695A.209B}; \cite{2013ApJ...763...74L};
\yearcite{2014AJ....148...37L}; \yearcite{2015AJ....149...93L}
\cite{2014AJ....147...45C}; \cite{2015AJ....149..197Z};
\cite{2015A&A...577A.146B}; \cite{2022A&A...668A.173G};
\cite{2022ApJ...924...66Y};
\cite{2023MNRAS.521.1908M}). \authorcite{2019MNRAS.485.2562H}
(\yearcite{2019MNRAS.485.2562H}; \yearcite{2022MNRAS.509..246H}) have
made a list of triple star system candidates, based on eclipse time
variation analysis of OGLE-IV eclipsing binaries. \tprim is not
included in triple star systems discovered by these surveys, since
\tprim is outside of the Kepler mission star field and the OGLE-IV
field. Many triple star systems have been discovered via eclipse time
variations from TESS (\cite{2020MNRAS.493.5005B};
\yearcite{2025arXiv251004565B}; \cite{2020MNRAS.498.6034M};
\yearcite{2024A&A...685A..43M}; \cite{2022MNRAS.513.4341R};
\yearcite{2024A&A...686A..27R}; \cite{2023MNRAS.522.3076Y};
\cite{2024A&A...690A.153M}; \cite{2024ApJ...974...25K};
\cite{2022MNRAS.511.4710E}). However, these studies have not found
\tprim.

\citet{2023MNRAS.526.2830C} have provided a list of triple star system
candidates by cross-matching the table ``\verb|nss_two_body_orbit|''
in Gaia DR3 with several databases of eclipsing binaries, such as
APASS (AAVSO Photometric All Sky Survey: \cite{2014AJ....148...81M}),
ASAS-SN (All-Sky Automated Survey for Supernovae:
\cite{2014ApJ...788...48S}; \cite{2022MNRAS.517.2190R}), GCVS (General
Catalog of Variable Stars: \cite{2017ARep...61...80S}), Kepler (Kepler
Eclipsing Binary Stars: \cite{2016AJ....151...68K}), TESS (TESS
Eclipsing Binary Stars: \cite{2022ApJS..258...16P}), and VSX (The
International Variable Star Index:
\cite{2006SASS...25...47W}). However, \citet{2023MNRAS.526.2830C} have
not included \tprim in the list of the triple star system
candidates. This is because \tprim is not listed in these
database. \citet{2024A&A...692A.247B} have also made a candidate list
of triple star systems from binaries with astrometric solutions in
Gaia DR3, or binaries categorized into \verb|Orbital| and
\verb|AstroSpectroSB1| by a data type \verb|nss_solution_type|.
Because \tprim has only a spectroscopic solution in Gaia DR3 (or
categorized into \verb|SB1|), \tprim is absent in the list.

\citet{2022ApJS..258...16P} have stated that it is difficult for TESS
to find eclipsing binaries with orbital periods of $\gtrsim 13$ days
due to its duty cycle. That should be the reason why TESS Eclipsing
Binary Stars do not include \tprim as an eclipsing binary as described
above. We can identify \tprim as an eclipsing binary, because we
concentrate \tprim, and know in advance that \tprim is a triple star
system.

\begin{figure}
 \begin{center}
   \includegraphics[width=8cm]{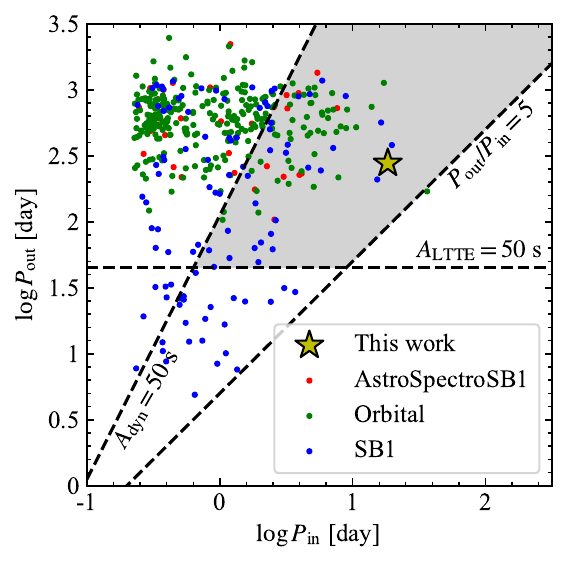}
 \end{center}
 \caption{Relation between inner and outer orbital periods of \tprim
   and triple star system candidates \citet{2023MNRAS.526.2830C}. The
   star sign indicates \tprim. Dots show the triple star system
   candidates, and their color codes are linked with Gaia NSS orbital
   solutions. The dashed lines indicate $\pout/\pin = 5$, $A_{\rm
     LTTE} = 50$ s, and $A_{\rm dyn} = 50$ s. When we calculate
   $A_{\rm LTTE}$ and $A_{\rm dyn}$, we set $\massp = \masss = \masst
   = 1\;\msun$, $\eout = 0.35$, $\iout = 60$ degrees, and $\argpr =
   90$ degrees. The shaded region indicates dynamically stable triple
   star systems which are easy to be discovered by the analysis of
   eclipse time variations \citep{2016MNRAS.455.4136B}.}
\label{fig:tripleparam}
\end{figure}

Figure \ref{fig:tripleparam} shows the relation between inner and
outer binary periods in triple star system candidates compiled by
\citet{2023MNRAS.526.2830C} based on Gaia DR3 and TESS. The star sign
indicates \tprim. We indicate a line of $\pout/\pin=5$, above which
triple star systems are dynamically stable
\citep{2001MNRAS.321..398M}. We notate amplitudes of eclipse time
variations due to the light-travel time effect and secular effect as
$A_{\rm LTTE}$ and $A_{\rm dyn}$, respectively (e.g.,
\cite{2015MNRAS.448..946B}). The amplitude of the light-travel time
effect is expressed as
\begin{align}
  \left( \frac{A_{\rm LTTE}}{{\rm day}} \right) \sim 1.1 \times
  10^{-4} &\left[ \frac{\gmas}{\msun} \right]^{1/3} \left(
  \frac{\pout}{{\rm day}} \right)^{2/3} \nonumber\\
  &\times \sqrt{1-\eout^2 \cos^2 \argpr},
\end{align}
where
\begin{align}
  \gmas = \frac{\massp^3}{(\massp+\masss+\masst)^2} \sin^3
  \iout. \label{eq:Massfunction2}
\end{align}
Note that Eq. (\ref{eq:Massfunction2}) is similar to
Eq. (\ref{eq:Massfunction}), however masses in numerators are
different. The amplitude of the secular effect is given by
\begin{align}
  A_{\rm dyn} = \frac{1}{2\pi}\frac{\massp}{\massp+\masss+\masst}
  \frac{\pin^2}{\pout} (1-\eout^2)^{3/2}.
\end{align}

The shaded region indicates dynamically stable triple star systems
which are easy to be discovered by the analysis of eclipse time
variations \citep{2016MNRAS.455.4136B}. Although \tprim is located in
the shaded region, we find \tprim not by the analysis of eclipse time
variations. We recognize that \tprim is a triple star system before
finding their eclipses. We use their eclipses solely to determine the
inner binary orbit. This means that we can find triple star systems
without eclipse time variations, combining low- and high-SNR
spectroscopic observations with Gaia DR3. Note that we will determine
inner orbits of triple systems without eclipses unlike \tprim, if we
will perform high SNR spectroscopy like our HDS observation several
times.

We can see in Figure \ref{fig:tripleparam} that \tprim has a small
outer-to-inner period ratio ($\sim 15$), compared with triple star
system candidates from Gaia DR3 \citep{2023MNRAS.526.2830C}. This is
true when we compare \tprim with triple star systems discovered from
the Kepler field (\cite{2016MNRAS.455.4136B};
\yearcite{2025A&A...695A.209B}). As for the outer eccentricity, inner
binary mass, and mass ratio of the primary star to the inner binary,
\tprim is similar to triple star systems found with eclipse time
variations by \citet{2025A&A...695A.209B}.

\section{Conclusions}
\label{sec:Conclusions}

We discover a triple star system \tprim, while searching for compact
binaries. We perform follow-up spectroscopic observations several
times with GAOES-RV, MALLS, and HDS, determine its outer orbit by RV
variation, and find the outer orbital period and eccentricity of $\sim
277.2$ days and $\sim 0.230$, respectively. We detect an inner binary
in \tprim with the SB3 spectral fit for a high-SNR spectroscopic
observation performed by Subaru HDS. We adopt the isochrone fitting,
and estimate the primary, secondary, and tertiary masses of $\sim
0.85$, $\sim 0.63$, and $\sim 0.61\;\msun$, respectively. The inner
orbit has the orbital period of $\sim 18.3$ days, and a small orbital
eccentricity ($\sim 0.01$), according to the combination of our
findings and the TESS data. Although \tprim has a relatively small
outer-to-inner period ratio ($\sim 15$), compared with triple star
systems discovered before, the other parameters are typical for triple
star systems.

In the last decade and a half, triple star systems like \tprim have
been discovered via eclipse time variations. On the other hand, we
find \tprim without adopting eclipse time variation. We only use
eclipses in order to determine the inner orbit of \tprim. We recognize
the presence of the inner binary before finding these eclipses. For
\tprim, we can determine the inner orbit by analyzing these eclipses
fortunately. Even unless \tprim indicated eclipses, we could determine
its inner orbit, performing high-SNR spectroscopic observations
several times. Our discovery implies the possibility that we can find
triple star systems without eclipse time variations, combining low-
and high-SNR spectroscopic observations with Gaia DR3, and the
upcoming Gaia DR4. Especially, such triple star systems will be found
in the course of searching for compact binaries like our survey.

Finally, we emphasize that a high-SNR ($\gtrsim 200$ per pixel)
spectroscopic observation with a 10-m class telescope is mandatory for
this discovery. Since the mass ratio of each inner binary component to
the primary star is small ($\sim 0.7$), absorption lines of each
binary component cannot be identified by a low-SNR ($\gtrsim 20$ per
pixel) spectroscopic observation. Moreover, a triple star system is
rare, in other words, distant and faint. We need to use a 10-m class
telescope, such as the Subaru telescope, in order to achieve high SNR
within reasonable time.

\begin{ack}
  This research is based on data collected at the 3.8-m Seimei
  telescope at the Okayama observatory, the 2-m NAYUTA telescope at
  the Nishi-Harima Astronomical Observatory (NHAO), and the Subaru
  Telescope. The 3.8-m Seimei telescope is operated by the National
  Astronomical Observatory of Japan, and Kyoto University. We are
  grateful to all the staff members at the Okayama observatory for
  their help during the observations. NHAO is operated by Center for
  Astronomy at University of Hyogo. We would like to express our
  gratitude to all the staff members at NHAO for their support during
  the observations.  The Subaru telescope is operated by the National
  Astronomical Observatory of Japan. We are honored and grateful for
  the opportunity of observing the Universe from Maunakea, which has
  the cultural, historical, and natural significance in Hawaii.

  The GAOES-RV project started as a collaboration between Gunma
  Astronomical Observatory and Institute of Science Tokyo. Based on
  the contract signed between the two parties, GAOES-RV is lent to
  Institute of Science Tokyo and operated at the Seimei Telescope.

  This work presents results from the European Space Agency (ESA)
  space mission Gaia. Gaia data are being processed by the Gaia Data
  Processing and Analysis Consortium (DPAC). Funding for the DPAC is
  provided by national institutions, in particular the institutions
  participating in the Gaia MultiLateral Agreement (MLA). The Gaia
  mission website is \url{https://www.cosmos.esa.int/gaia}. The Gaia
  archive website is \url{https://archives.esac.esa.int/gaia}.
  
  This research has made use of the VizieR catalogue access tool, CDS,
  Strasbourg, France \citep{10.26093/cds/vizier}. The original
  description of the VizieR service was published in
  \citet{vizier2000}. This research has used data, tools or materials
  developed as part of the EXPLORE project that has received funding
  from the European Union’s Horizon 2020 research and innovation
  programme under grant agreement No 101004214.

  We acknowledge the use of TESS High Level Science Products (HLSP)
  produced by the Quick-Look Pipeline (QLP) at the TESS Science Office
  at MIT, which are publicly available from the Mikulski Archive for
  Space Telescopes (MAST). Funding for the TESS mission is provided by
  NASA's Science Mission directorate.
\end{ack}

\section*{Funding}
This research was supported by Grants-in-Aid for Scientific Research,
24K07040 and 25K01035 (A.T.) from the Japan Society for the Promotion
of Science.

\section*{Data availability} 

The data underlying this article are available publicly.





\end{document}